\documentclass[apj,revtex4]{emulateapj}
\usepackage{textcomp}
\usepackage{epsfig, epstopdf}
\usepackage{longtable}
\usepackage{lscape}
\usepackage{amssymb,amsmath}

\newcommand{\cf}{{\ifmmode{C_{\rm f}}\else{$C_{\rm f}$}\fi}}
\newcommand{\zem}{{\ifmmode{z_{\rm em}}\else{$z_{\rm em}$}\fi}}
\newcommand{\zabs}{{\ifmmode{z_{\rm abs}}\else{$z_{\rm abs}$}\fi}}
\newcommand{\zs}{{\ifmmode{z_{\rm s}}\else{$z_{\rm s}$}\fi}}
\newcommand{\zl}{{\ifmmode{z_{\rm l}}\else{$z_{\rm l}$}\fi}}
\newcommand{\kms}{{\ifmmode{{\rm km~s}^{-1}}\else{km~s$^{-1}$}\fi}}
\newcommand{\vej}{{\ifmmode{v_{\rm ej}}\else{$v_{\rm ej}$}\fi}}
\newcommand{\vrot}{{\ifmmode{v_{\rm rot}}\else{$v_{\rm rot}$}\fi}}
\newcommand{\cm}{{\ifmmode{{\rm cm}^{-1}}\else{cm$^{-1}$}\fi}}
\newcommand{\cmm}{{\ifmmode{{\rm cm}^{-2}}\else{cm$^{-2}$}\fi}}
\newcommand{\cmmm}{{\ifmmode{{\rm cm}^{-3}}\else{cm$^{-3}$}\fi}}

\newcounter{species} 
\def\ion#1#2{\setcounter{species}{#2}#1$\;${\scriptsize\Roman{species}}\relax}

\shorttitle{NAL acceleration}
\shortauthors{Misawa et al.}

\begin{document}

\title{Direct Measurement of Quasar Outflow Wind
  Acceleration\altaffilmark{1,2,3}}

\footnotetext[1]{Based on data collected at Subaru Telescope, which is
  operated by the National Astronomical Observatory of
  Japan.}
\footnotetext[2]{Data presented herein were obtained at the W.M. Keck
  Observatory, which is operated as a scientific partnership among the
  California Institute of Technology, the University of California and
  the National Aeronautics and Space Administration. The Observatory
  was made possible by the generous financial support of the W.M. Keck
  Foundation.}
\footnotetext[3]{Based on observations obtained at the European
  Southern Observatory at La Silla, Chile in programs 65.O-0474(A) and
  079.B-0469(A)}

\author{Toru Misawa$^4$, 
        Michael Eracleous$^{5,6}$, 
        Jane C. Charlton$^5$, 
    and Nobunari Kashikawa$^7$
}

\affil{
$^4$School of General Education, Shinshu University, 3-1-1 Asahi,
  Matsumoto, Nagano 390-8621, Japan; misawatr@shinshu-u.ac.jp \\
$^5$Department of Astronomy \& Astrophysics, Pennsylvania State
  University, 525 Davey Lab, University Park, PA 16802 \\
$^6$Institute for Gravitation and the Cosmos, Pennsylvania State
  University, 525 Davey Lab, University Park, PA 16802 \\
$^7$Department of Astronomy, School of Science, The University of
  Tokyo, 7-3-1 Hongo, Bunkyo-ku, Tokyo 113-0033, Japan\\
}

\begin{abstract}
We search for velocity changes (i.e., acceleration/deceleration) of
{\it narrow} absorption lines (NALs) that are {\it intrinsic} to the
quasars, using spectra of 6 bright quasars that have been observed
more than once with 8--10m class telescopes.  While variations in line
strength and profile are frequently reported (especially in broader
absorption lines), definitive evidence for velocity {\it shifts} has
not been found with only a few exceptions.  Direct velocity shift
measurements are valuable constraints on the acceleration mechanisms.
In this study, we determine velocity shifts by comparing the
absorption profiles of NALs at two epochs separated by more than
10~years in the observed frame, using the cross-correlation function
method and we estimate the uncertainties using Monte Carlo
simulations.  We do not detect any significant shifts but we obtain
3$\sigma$ upper limits on the acceleration of intrinsic NALs (compared
to intervening NALs in same quasars) of
$\sim$0.7~km~s$^{-1}$~yr$^{-1}$ ($\sim$0.002~cm~s$^{-2}$).  We discuss
possible scenarios for non-detection of NAL acceleration/deceleration
and examine resulting constraints on the physical conditions in
accretion disk wind.
\end{abstract}


\section{Introduction}
The signatures of outflowing winds from accretion disks are observed
in the spectra of $\sim$50\%\ of quasars and active galactic nuclei
\citep[hereafter AGNs;][]{ham12}.  The winds play an important role in
providing energy and momentum feedback to the interstellar medium
(ISM) of their host galaxies and the inter-galactic medium (IGM), and
by transporting heavy elements out of the host galaxies
\citep[e.g.,][]{spr05}.  The blueshifted near-UV absorption lines
(e.g., the \ion{C}{4}, \ion{Si}{4}, and \ion{N}{5} doublets) that are
the observational signature of outflowing winds, are usually
classified into three categories according to their widths: broad
absorption lines (BALs) with FWHM $\geq$ 2,000~km/s, narrow absorption
lines (NALs) with FWHM $\leq$ 500~km/s, and an intermediate subclass
(mini-BALs).  The difference in line width may depend on the
inclination angle of our line of sight relative to the axis of the
wind (an orientation scenario; e.g., \citealt{wey91}), although
evolutionary scenarios have also been discussed
\citep[e.g.,][]{lip06}; see summary in \citet{bru12} and references
therein.

The physical conditions of quasar winds have been studied through
variability of the absorption lines since almost all BALs and
mini-BALs vary over a few years in the quasar's rest-frame
\citep[e.g.,][]{gib08,mis14}. Variations are usually seen in the
strengths and profiles of (mini-)BAL troughs \citep[e.g.,][]{cap12}. The
probability of observing variations increases with the time interval
between observations, approaching unity for an interval of a few years
\citep[e.g.,][]{cap11}. In extreme cases, BAL profiles appear or
disappear completely \citep[e.g.,][]{viv12,fil12}.  Possible origins
of such variations include (a) motion of the absorbing gas
parcels across our line of sight \citep[e.g.,][]{ham08} and (b)
changes in the ionization state of the absorber
\citep[e.g.,][]{mis07b}.

{\it Velocity shifts} of absorption lines from winds have not been
studied except in a few cases for BALs and mini-BALs
\citep[e.g.,][]{jos18,gri16,jos14}, despite the fact that acceleration
is a key physical property of the gas in quasar winds. A direct
detection of velocity shifts would provide valuable constraints on
acceleration mechanisms.  Three different processes have been proposed
for accelerating winds: radiation pressure \citep[absorption of line
  photons; e.g.,][]{mur95}, magnetocentrifugal forces
\citep[e.g.,][]{eve05}, and thermal pressure \citep[e.g.,][]{bal93}.
Among them, radiation pressure must contribute significantly at least
to the acceleration of X-ray absorbers with large ejection velocity
(Ultra-Fast Outflows or UFOs), because their total kinetic energy,
$\dot{E}_k = \frac{1}{2} \dot{M} v^2$ (where $\dot{M}$ is the mass
ejection rate) is well correlated with the bolometric luminosity
($L_{bol}$) of the quasars \citep[e.g.,][]{tom13}.  However, it is
still not clear whether the radiative force is also the main source of
acceleration for absorbers detected through {\it UV lines} whose
distances from the central engine may be much larger than those of
absorbers detected via X-ray lines. Studies of velocity shifts using
(mini-)BALs may be complicated by the following: (1) the absorption
lines sometimes suffer from self-blending (i.e., the blue and red
components of a doublet are blended with each other), and (2) a change
in line profile could be mistaken as a velocity shift.

In contrast to previous studies that focused on BALs and mini-BALs, we
monitor here the velocities of {\it intrinsic NALs}, {\it for the
  first time}. In the case of NALs, small velocity shifts can be
detected very easily because of their small widths. We describe target
selection in \S2, the observations and data reduction in \S3, and the
data analysis in \S4.  The results and discussion are presented in \S5
and \S6.  We close by summarizing our results in \S7.  We use a
cosmology with $H_{0}$=70 \kms~Mpc$^{-1}$, $\Omega_{m}$=0.3, and
$\Omega_{\Lambda}$=0.7 throughout the paper.

\section{Sample Selection}
We assembled a sample of six quasars based on two criteria: a) they
have at least one \ion{C}{4}, \ion{Si}{4}, or \ion{N}{5} NAL in their
spectra that has already been confirmed to be physically associated
with the quasar, and b) there are multiple high-resolution spectra
($R$ $\sim$ 40,000) of each quasar that were already available or
would soon be available through upcoming observing programs over a
time interval of more than 10~years in the observed frame.

We selected the target quasars with ``reliable'' (class A) and
``possible'' (class B) intrinsic NALs from the catalog of
\citet{mis07a}. The catalog was constructed after examining the NALs
in 37 quasar spectra and selecting those that are intrinsic based on
partial coverage analysis.  Two of the quasars we selected,
HE0130--4021 and HS1700+6416, had previously been observed more than
once with 10~m class telescopes (i.e., the Keck and Subaru Telescopes,
and the Very Large Telescope, hereafter VLT) with some time intervals
that exceed 10 years in the observed frame \citep[see details
  in][]{mis14}.  We also chose four additional quasars, Q0450--1310,
HE0940--1050, HE1009+2956, and Q1107+4847, which we re-observed with
the Subaru Telescope. These four quasars are known to have at least
one {\it intrinsic} NAL and their oldest spectra were taken more than
10 years ago. They are accessible from the Subaru Telescope, and are
bright enough ($V$ $\leq$ 18.2) so that spectra with a signal-to-noise
(S/N) ratio of $\geq$20~pix$^{-1}$ could be obtained with $\sim$2 hour
exposures.

Table~\ref{tab:qso} summarizes the properties of our sample quasars
including coordinates, emission redshift, optical magnitude,
bolometric luminosity, and radio-loudness.

\section{New Observations}
We observed the four quasars noted in the previous section with the
Subaru Telescope and High Dispersion Spectrograph (HDS) on 2016
January 27.  The weather conditions were very good throughout the
night with a typical seeing of $\sim$$0.\!\!^{\prime\prime}6$.  The
time delays from the first observations of those quasars are
$\sim$15--20 years in the observed frame.  We acquired high-resolution
spectra ($R$ $\sim$ 45,000) with a slit width of
$0.\!\!^{\prime\prime}8$ and adopted bins of 2 pixels in both the
spatial and dispersion directions (i.e., $\sim$0.03~\AA\ per pixel) to
increase the S/N ratio.  To cover as many NALs as possible, we used
standard setups, Std-Ya for Q0450--1310 and HE1009+2956 and Std-Yc for
HE0940--1050 and Q1107+4847. These cover a wavelength range of
4030--4800~\AA\ or 4420--5660~\AA\ on the blue CCD and
4940--5660~\AA\ or 5860--7050~\AA\ on the red CCD, respectively.  We
took four 1800s exposures for each quasar (i.e., a total integration
time of 7200s), except for Q0450--1310 whose total integration time
was 10800s (1800s $\times$ 6).

We reduced the data in a standard manner with the
IRAF\footnote[8]{IRAF is distributed by the National Optical Astronomy
  Observatories, which are operated by the Association of Universities
  for Research in Astronomy, Inc., under cooperative agreement with
  the National Science Foundation.} software package.  Wavelength
calibration was performed with the help of spectra of a Th-Ar lamp.
We fitted the {\it effective} continuum, which also includes a
substantial contribution from broad emission lines, with a third-order
cubic spline function to normalize the spectra.  We did not adjust the
wavelength resolution of the new spectra to match that of the old
spectra, because the typical line width of absorption components is
large enough ($b$ $\geq$ 10~\kms; \citealt{mis07a}) that they are
fully resolved.  The final S/N ratio around $\sim$4900\AA\ is
$\sim$15--47 per pixel for each quasar.  The log of our new and old
observations is given in Table~\ref{tab:qso}, including observation
date, spectral resolution, total integration time, and S/N ratio per
pixel. In the same table we also give the rest-frame time interval
between the observations.

\section{Cross-Correlation Analysis}
\label{sec:crosscorr}
We measured velocity shifts by comparing NAL profiles in the first and
second epochs, using the cross-correlation function method that was
originally adopted for BALs in \citet{gri16}.

\begin{figure*}
 \begin{center}
  \includegraphics[width=8cm,angle=270]{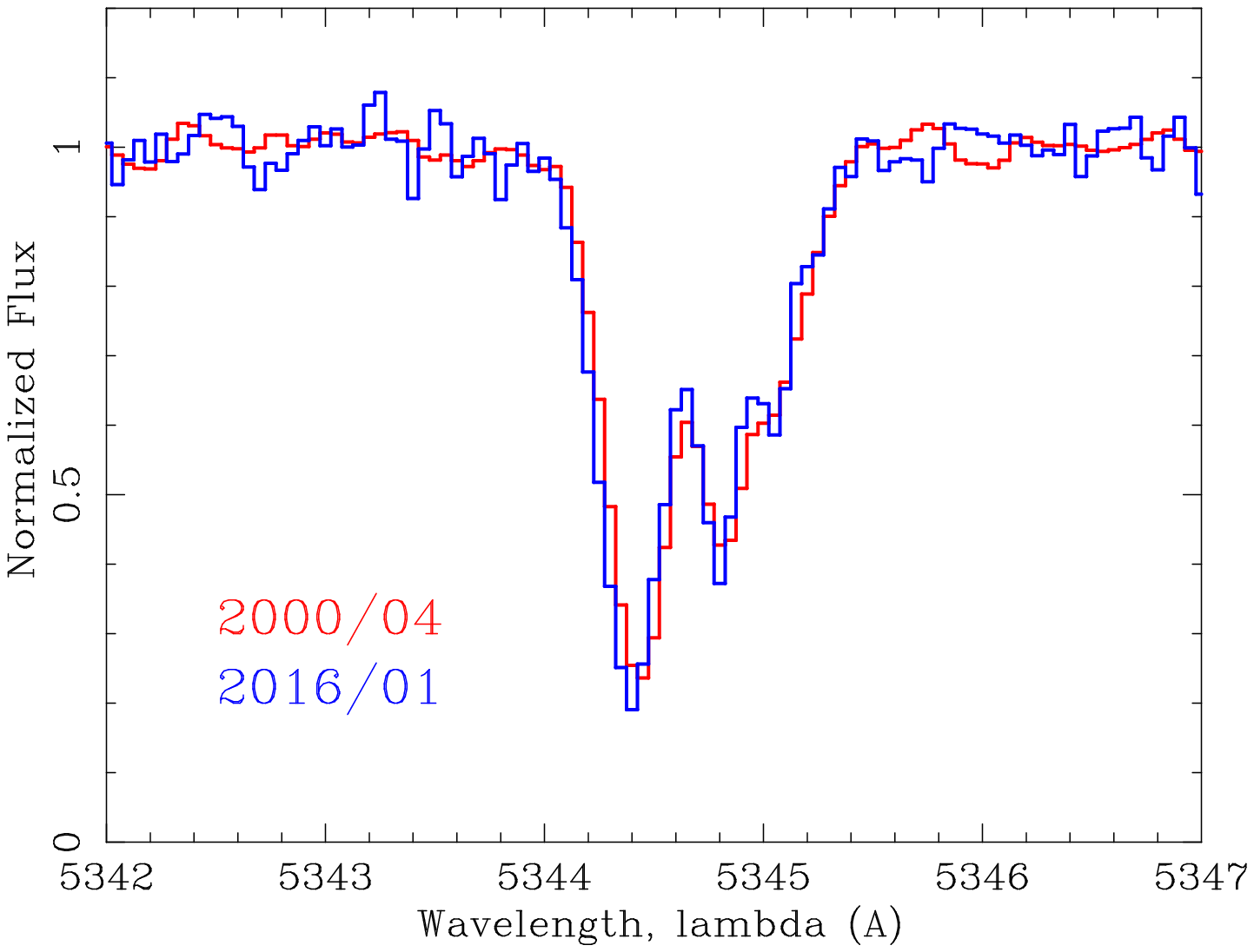}
 \end{center}
 \vspace{1cm}
 \begin{center}
  \plottwo{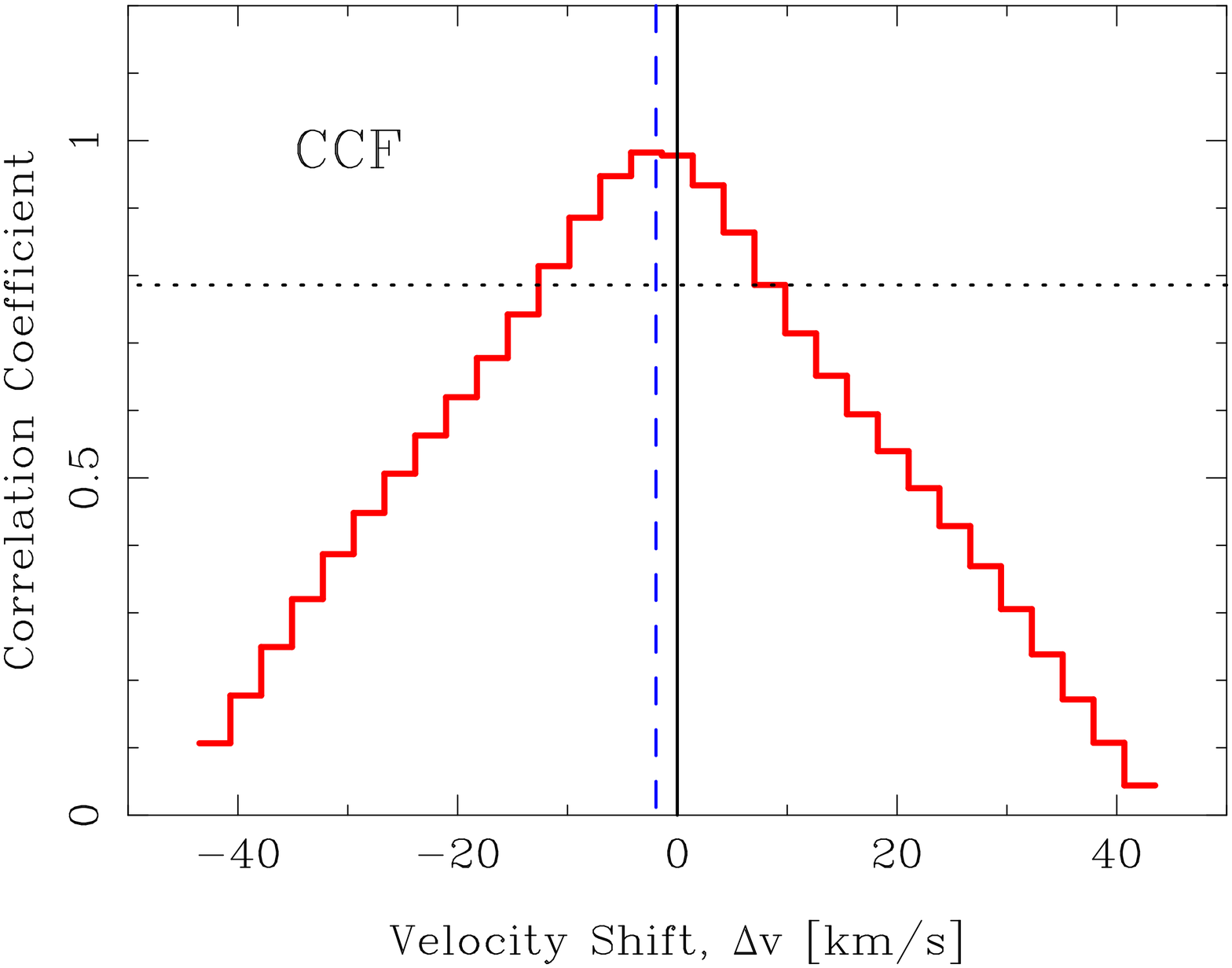}{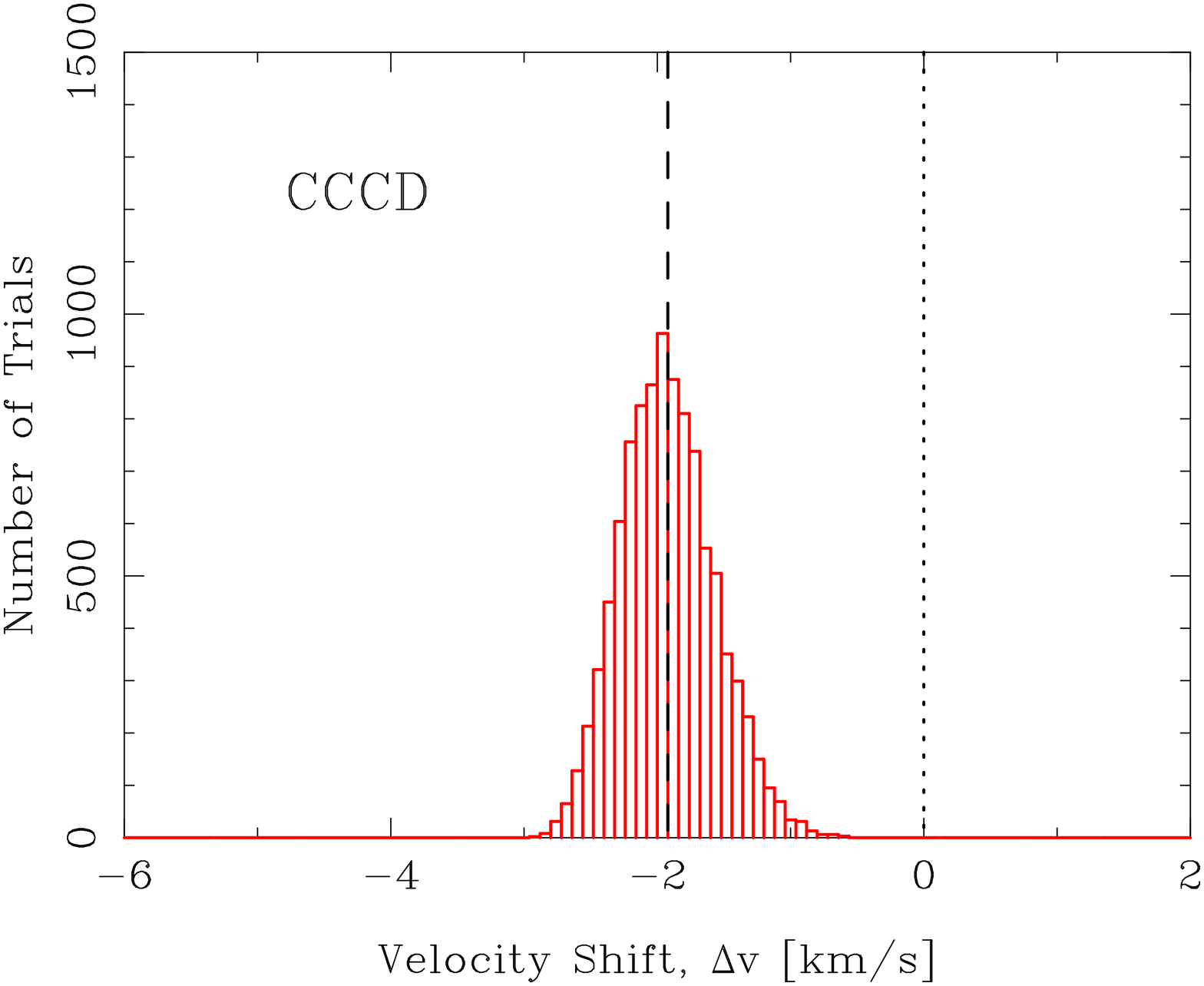}
 \end{center}
\caption{The top panel shows absorption profile of the blue-component
  of the class-B \ion{Si}{4} NAL (i.e., \ion{Si}{4}~$\lambda$1394) at
  \zabs = 2.8347 in HE0940--1050. The spectra were taken on 2000 April
  3 and 2016 January 27. The Cross-Correlation Function (CCF) and
  Cross-Correlation Centroid Distribution (CCCD) for the
  \ion{Si}{4}$\lambda$1394 line are shown in the bottom left and right
  panels, respectively, as an example.\label{fig:ccf_cccd}}
\end{figure*}

We first isolated a spectral region around each NAL with
$\sim$2\AA\ margins on both sides\footnote[9]{We require at least
  $\pm$1~\AA\ on both sides even if the $\pm$2\AA\ regions are
  partially affected by data defects or line blending.}, and resampled
it to bins of 0.05~\AA, which is the typical dispersion (pixel scale)
of all spectra in our sample (see Figure~\ref{fig:velplot}).  Next, we
normalized the clipped spectrum using a low-order spline function
avoiding absorption regions.  Finally, we used the spectra from the
first and second epochs to calculate Pearson's cross-correlation
coefficient,
\begin{equation}
r =
\frac{\sum(f_1(\lambda)-\overline{f_1(\lambda)})(f_2(\lambda)-\overline{f_2(\lambda)})}{\sqrt{\sum(f_1(\lambda)-\overline{f_1(\lambda)})^2}\sqrt{\sum(f_2(\lambda)-\overline{f_2(\lambda)})^2}},
\end{equation}
where $f_1(\lambda)$ and $f_2(\lambda)$ are the normalized flux
densities in the first and second epochs and $\overline{f_1(\lambda)}$
and $\overline{f_2(\lambda)}$ are their average values.  We repeated
the calculation of $r$ after shifting one spectrum from $-$15~pixel to
$+$15~pixel in steps of 0.05~\AA\ to build the Cross Correlation
Function (CCF). From the CCF we measured the velocity shift
corresponding to the CCF centroid, $\Delta v$(CCF), by finding the
weighted center of the CCF using the portion corresponding to $r > 0.8
r_{\rm peak}$, where $r_{\rm peak}$ is the peak value of the CCF.
We performed this analysis separately for the blue and
red members of doublets (i.e., \ion{C}{4}$\lambda\lambda$1548,1551,
\ion{Si}{4}$\lambda\lambda$1394,1403, and
\ion{N}{5}$\lambda\lambda$1239,1243), if both were detected.  As an
example, the CCF for the class-B \ion{Si}{4} NAL at \zabs\ = 2.8347 for
HE0940--1050 is shown in Figure~\ref{fig:ccf_cccd}.  The $\Delta
v$(CCF) values for all the NALs are summarized in Table~\ref{tab:nal}.

To determine the uncertainties, we performed a Monte Carlo simulation
using synthetic spectra in which the flux density of each pixel was
perturbed by a random amount drawn from a Gaussian distribution with a
mean equal to the measured flux density and a standard deviation equal
to the uncertainty.  We constructed 100 synthetic spectra for each
epoch, and repeated the CCF calculations 10,000 times using all
combinations of the artificial spectra in both epochs (i.e.,
100$\times$100).  We measured the average of the CCF centroid
(Cross-Correlation Centroid Distribution; $\Delta v$(CCCD)) and its
1$\sigma$ uncertainty for each NAL, and we summarize these in
Table~\ref{tab:nal}.

To assess the performance of the CCF and CCCD methods used above,
we carried out the following tests

\begin{figure*}
 \begin{center}
  \includegraphics[width=10cm,angle=0]{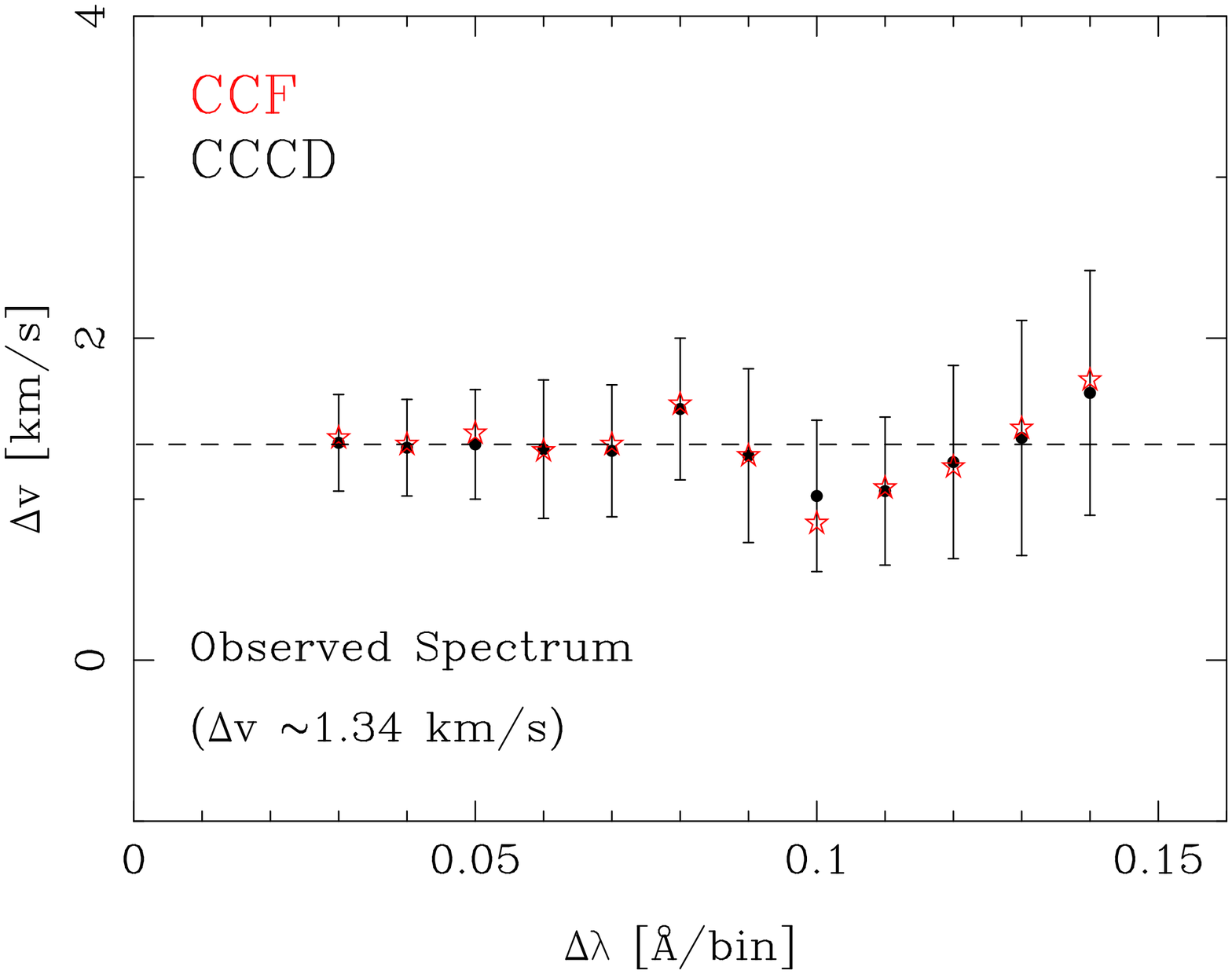}
 \end{center}
 \vspace{1cm}
 \begin{center}
  \includegraphics[width=10cm,angle=0]{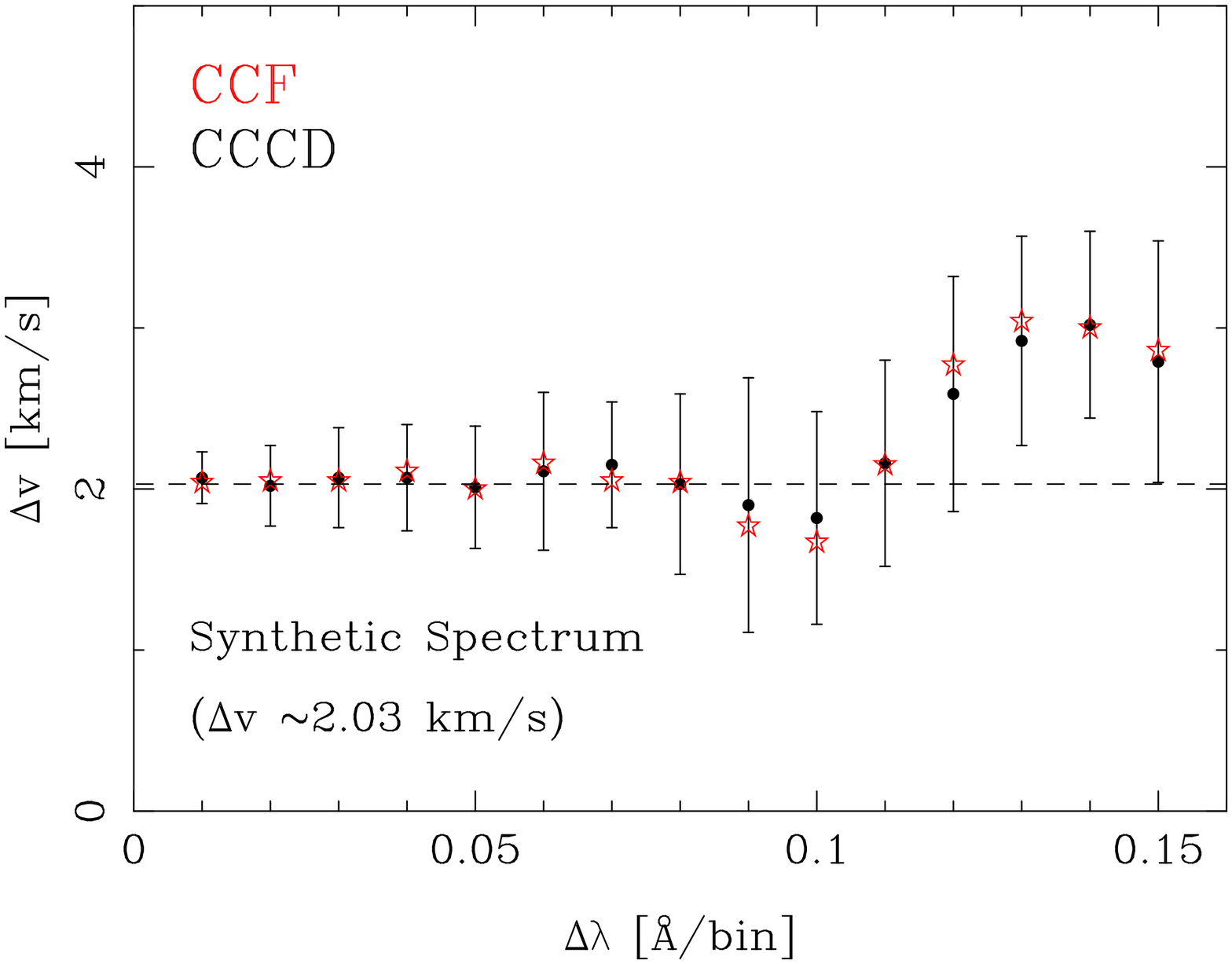}
 \end{center}
\caption{Results of tests of the CCF method used in our analysis. The
  two panels show the measured velocity shift between the two spectra
  in a pair as a function of dispersion (i.e., size of wavelength
  bins). An example of results from an observed spectrum, the
  \ion{C}{4}~$\lambda$1548 line at \zabs\ = 2.2316 toward
  HE0130$-$4021, is shown in the top panel. An example based on a
  synthetic spectrum of a similar \ion{C}{4} line with a
  2.03~\kms\ shift is shown in the bottom panel.  See details in
  \S~\ref{sec:crosscorr} of the text.\label{fig:test1}}
\end{figure*}

\begin{figure*}
 \begin{center}
  \includegraphics[width=10cm,angle=0]{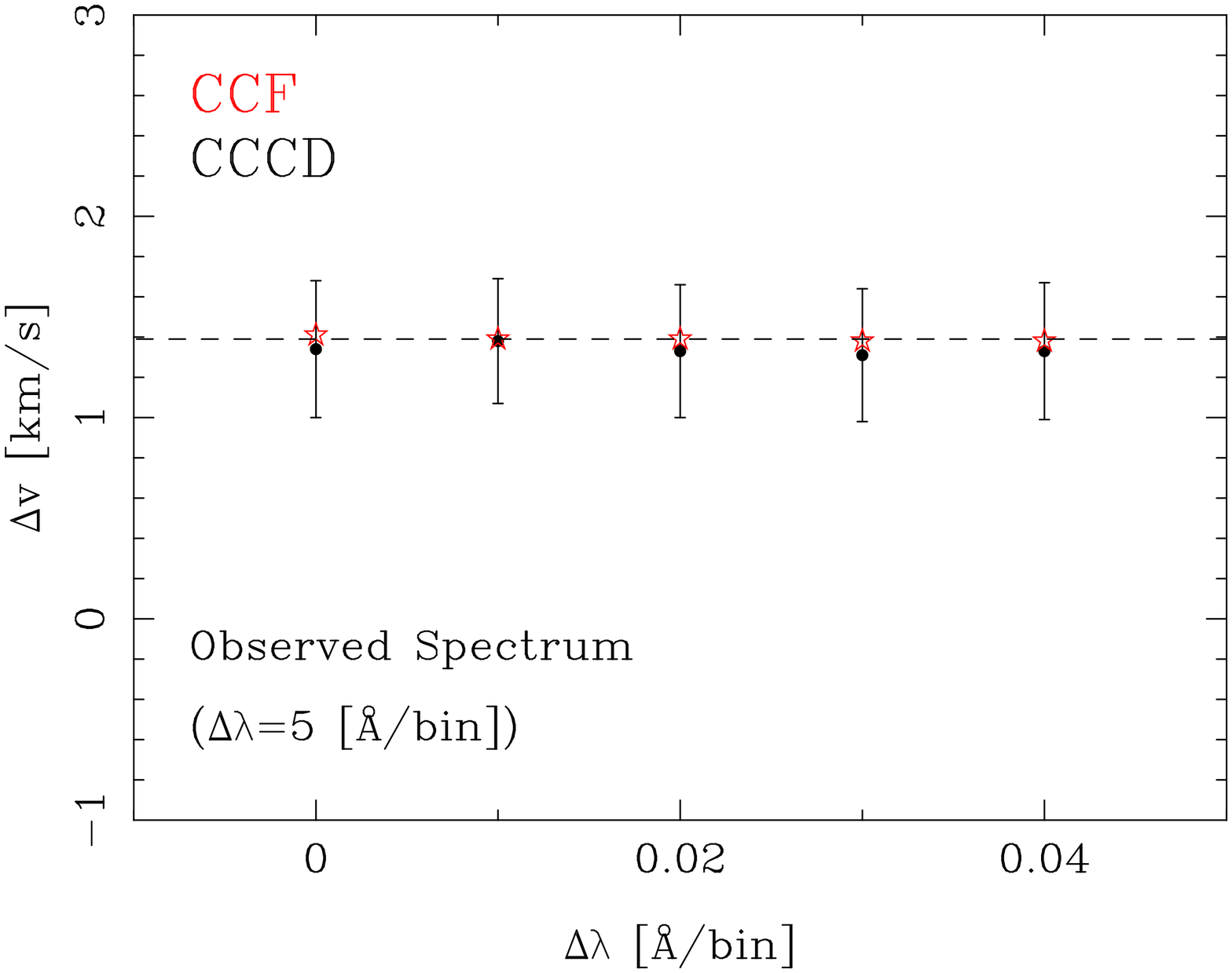}
 \end{center}
 \vspace{1cm}
 \begin{center}
  \includegraphics[width=10cm,angle=0]{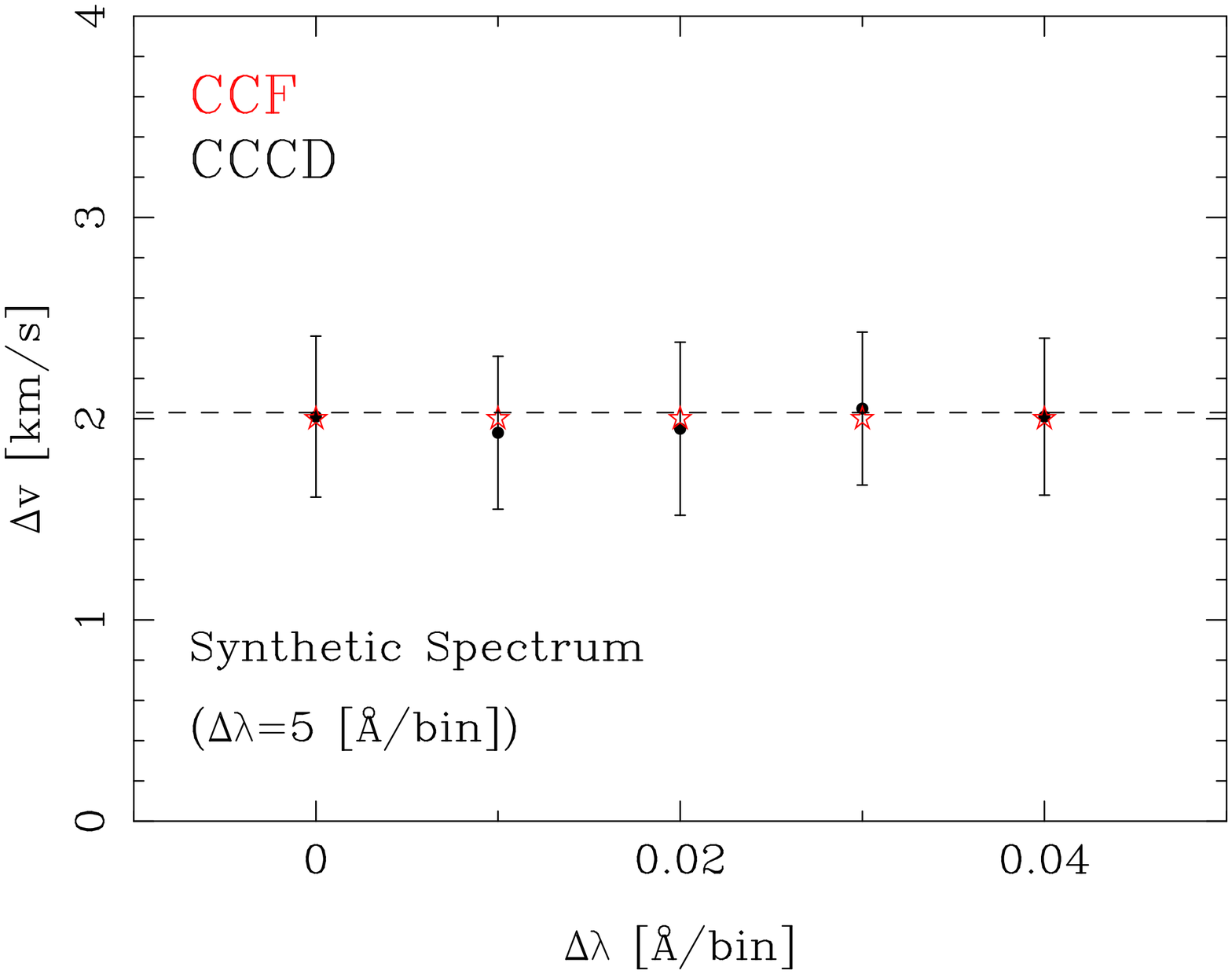}
 \end{center}
\caption{Results of tests of the CCF method used in our analysis. The
  two panels show the measured velocity shift between the two spectra
  in a pair as a function of starting wavelength. An example of
  results from an observed spectrum, the \ion{C}{4}~$\lambda$1548 line
  at \zabs\ = 2.2316 toward HE0130$-$4021, is shown in the top panel.
  An example based on a synthetic spectrum of a similar \ion{C}{4}
  line with a 2.03~\kms\ shift in the bottom panel.  See details in
  \S~\ref{sec:crosscorr} of the text.\label{fig:test2}}
\end{figure*}

\begin{enumerate}

\item
  We investigated whether the sampling of the spectra led to
  systematic errors in the shift determination.  We synthesized two
  normalized spectra corresponding to those in the first and second
  epochs, in which we added a single component of a
  \ion{C}{4}~$\lambda$1548 line at \zabs = 2.2367 and 2.236722,
  respectively.  Both spectra have a wavelength range of $\lambda$ =
  5007.5--5014.5\AA.  The velocity offset between the \ion{C}{4}
  lines\footnote[10]{This is not due to the cosmic expansion, but
    calculated by the relativistic Doppler formula,
\begin{equation}
  {\Delta v\over c} =
  -\frac{(1+z_1)^2-(1+z_2)^2}{(1+z_1)^2+(1+z_2)^2},
\end{equation}
where $z_1$ and $z_2$ are apparent redshift of the \ion{C}{4} absorber
at the first and second epochs, respectively.} is 2.03~\kms.  We used
  common line parameters except for redshift: column density and
  Doppler parameter of $\log (N_{\rm C\;IV}/\cmm) = 13.5$ and $b =
  12.0$~\kms\ (i.e., typical values of our NAL sample), and a coverage
  fraction\footnote[11]{The fraction of photons from the background
    light source(s) that pass through the absorber
    \citep[e.g.,][]{wam95,bar97}} of \cf\ = 1.0 (i.e., a full
  coverage).  We also assumed a wavelength resolution of $R$ = 45,000
  (a typical value of our data), a dispersion of $\Delta \lambda$ =
  0.01\AA\ per pixel, and finally added a noise corresponding to an
  uncertainty of 0.033 in each pixel (i.e., S/N ratio $\sim$30
  pixel$^{-1}$).

First, we constructed synthetic spectra for a range of different
dispersions, from $\Delta \lambda$ = 0.01~\AA~pixel$^{-1}$ to
0.15~\AA~pixel$^{-1}$ in steps of 0.01\AA~pixel$^{-1}$, and used them
to perform the CCF and CCCD analyses and evaluate the shift between
them.  We also performed the same test on an observed spectrum, the
\ion{C}{4}~$\lambda$1548 line at \zabs\ = 2.2316 for HE0130--4021, for
which we found a velocity shift of $\Delta v$ = 1.34~\kms\ based on
the default dispersion (i.e., $\Delta v$ = 0.05\AA).  The results for
the {\it synthetic} and the {\it observed} spectra are shown in
Figure~\ref{fig:test1} as a function of dispersion.  Both results are
obviously affected by the wavelength bin size. The shift, $\Delta v$,
first decreases slightly with increasing $\Delta \lambda$ up to
0.1~\AA~pixel$^{-1}$ then decreases slightly around $\Delta\lambda$
$\sim$ 0.12~\AA~pixel$^{-1}$ and increases again towards
$\Delta\lambda$ $\sim$ 0.15~\AA~pixel$^{-1}$. Certainly, this pattern
should also depend on the spectral resolution, however, in our case
(i.e., $R$$ \sim$45,000), any systematic error in the shift would be
negligible since we adopt a dispersion of $\Delta\lambda <
0.07$~\AA~pixel$^{-1}$ or smaller.

We then carried out the same test while changing the starting
wavelength for sampling by one-fifth of the bin size (i.e., by
0.01~\AA). The results are shown in Figure~\ref{fig:test2}, which
illustrates that there the is no obvious systematic error.

\item
We compared the results from the above methods with shift measurements
made with the ``$\chi^2$ cross-correlation'' method of \citet[][see
  their {\S}6.1]{era12}. We also carried out simulations as described
in \citet{era12} to determine the smallest shifts measurable with the
available data and verify the reliability of the uncertainties.  We
found the results of the two independent methods of shift measurement
to be in very good agreement with each other.

\end{enumerate}

From the above tests, we conclude that the CCF and CCCD methods, as we
implement them here, are robust and are not affected by the wavelength
binning scheme that we adopted. Henceforth, we adopt the results of
the CCF and CCCD analysis, as summarized in Table~\ref{tab:nal}, and
we discuss them below.

We found the distributions of the $\Delta v$(CCF) and $\Delta v$(CCCD)
to be systematically shifted from zero (see Figure~\ref{fig:cccd}),
which we attribute to a small, systematic {\it linear offset} of the
spectra in each pair resulting from an uncertainty of wavelength
calibration using Th-Ar spectra. To quantify this systematic effect,
we measured the velocity shift of intrinsic (i.e., class-A and B) NALs
compared to intervening (i.e., class-C) NALs between the two epochs
for each quasar, instead of using the values of $\Delta v$(CCCD)
obtained by comparing directly the intrinsic NALs from the two epochs.

We also noticed that there is a systematic {\it non-linear distortion}
in relative wavelength between spectra at different epochs (e.g.,
\citealt{gri10,eva14}). For example, the $\Delta v$(CCCD) values for
the blue and red members of the \ion{C}{4} NAL at \zabs\ = 1.6968
toward Q0450--1310 (i.e., $-$4.26$\pm$1.40 and $-$0.91$\pm$2.53) do
not match, although they should be the same.  To evaluate the
distortion error\footnote[12]{Here, the distortion error should depend
  not only on wavelength calibration but on data quality i.e., the S/N
  ratio.}, we compared all pairs of intervening (class-C) lines and
examined the distribution of velocity shift differences (i.e.,
$|\Delta v|_{ij}$ = $|\Delta v{\rm (CCCD)}_i - \Delta v{\rm
  (CCCD)}_j|$ between {\it i}-th and {\it j}-th lines in a same
spectrum) and calculate their root mean square (r.m.s.) values as
summarized in Table~\ref{tab:rms}.  For example, we detect 14 class-C
lines (here, we double-count lines if both blue and red members of a
doublet are detected) in HE0940--1050, and calculate $|\Delta v|$ for
91 pairs (i.e., $_{14}C_2$) in total.  The velocity difference and its
{\it total} uncertainty, $\Delta v$ and $\sigma(\Delta v)$, for each
line, in which we add errors from the CCCD analysis and the non-linear
distortion in quadrature, are included in Table~\ref{tab:nal}.

Our sample quasars have multiple NALs in their spectra (from 3 NALs in
HE1009+2956 to 10 NALs in HE0940--1050). Therefore, we also determined
a weighted mean value of $\Delta v$ and its uncertainty for all the
class-A+B NALs in each quasar and, separately, all the class-C NALs in
each quasar using the following equations,
\begin{equation}
\Delta v = \frac{\sum_{\rm i=1}^{\rm n}\left(\frac{1}{\sigma(\Delta
    v_i)^2}\right)\Delta v_i}{\sum_{\rm i=1}^{\rm
      n}\left(\frac{1}{\sigma(\Delta v_i)^2}\right)} \; ,
\label{eqn:delv}
\end{equation}
and 
\begin{equation}
\sigma(\Delta v) = \sqrt{\frac{\sum_{\rm i=1}^{\rm
      n}\left(\frac{1}{\sigma(\Delta v_i)^2}\right)\left(\Delta v_{\rm
      i} - \overline{\Delta v}\right)^{2}}{(n-1)\sum_{\rm i=1}^{\rm
      n}\left(\frac{1}{\sigma(\Delta v_i)^2}\right)}} \; ,
\label{eqn:sigv}
\end{equation}
where $n$ is the number of absorption lines in a quasar and $\Delta
v_i$ and $\sigma(\Delta v_i)$ are the velocity shift and its
uncertainty between the observations for the $i$th line of a given
quasars, and $\overline{\Delta v}$ is their average. (i.e., $\sum_{\rm
  i=1}^{\rm n}\Delta v_i/n$). If $n < 2$, as in the class-A/B NALs in
HE1009+2956 and Q1107+4847, we take only the statistical error into
consideration and, instead of using equation \ref{eqn:sigv}, we
calculate $\sigma(\Delta v)$ as
\begin{equation}
\sigma(\Delta v) = \sqrt{\frac{1}{\sum_{\rm i=1}^{\rm
      n}\left(\frac{1}{\sigma(\Delta v_i)^2}\right)}} \; .
\label{eqn:sigv_alt}
\end{equation}
We also perform the same calculation using only {\it reliable} NALs
after removing those that i) show line-locking, ii) are partially
affected by line blending or have a data defect within 2~\AA\ of a
NAL, or iii) show $>$4$\sigma$ variability in their equivalent width
or line profile.  The resulting average $\Delta v$ and its
uncertainty, $\sigma(\Delta v)$, for all the class-A+B and the class-C
NALs in the quasars of our sample are shown in Figure~\ref{fig:cccd}
and tabulated in Table~\ref{tab:wa}.

\section{Results}
We applied the CCF and the CCCD analyses to 56
lines\footnote[13]{Here, we count the two lines of a doublet
  separately, if they are both clearly detected without any line
  blending or data defects.} in 40 NALs detected in six optically
bright quasars.  For the two NALs that exhibit
line-locking\footnote[14]{The red member of a doublet is aligned with
  the blue member of the following doublet. This is one of the
  indicators that a NAL is physically associated with a quasar
  \citep{ara96}.}, we applied the analyses to the blue and red members
of the doublets simultaneously.  The velocity plots of the NALs are
shown in Figure~\ref{fig:velplot}, except for the \ion{C}{4} NAL at
\zabs\ = 2.7243 in Q1107+4847 because it is severely affected by data
defect in the second spectrum.

As an example, we show the absorption profile of the blue member of the
class-B \ion{Si}{4} NAL (i.e., \ion{Si}{4}~$\lambda$1394) at \zabs =
2.8347 in HE0940--1050, and the results of the CCF and CCCD analyses
in Figure~\ref{fig:ccf_cccd}.  Below we describe in detail the results
for each quasar.  The results of the CCF and the CCCD analyses for
each NAL and each quasar are summarized in Table~\ref{tab:nal} and
Table~\ref{tab:wa}. The weighted averages of $\Delta v$ for class-A+B
and C NALs are also shown in Figures~\ref{fig:cccd}.

\begin{description}

\item[HE0130--4021] \citet{mis07a} detected six NALs (4 \ion{C}{4}, 1
  \ion{Si}{4}, and 1 \ion{N}{5}), of which three are classified as
  class-A/B NALs.  We observed all of these twice with a time interval
  of $\Delta t_{\rm obs}$ $\sim$11.7~yrs in the observed frame,
  corresponding to $\Delta t_{\rm rest}$ $\sim$ 2.90~yrs in the quasar
  rest frame.  Among the 12 blue/red lines of the six NALs, seven are
  detected without being affected by data defect or other unrelated
  lines.  The wavelength margin is smaller than 2\AA\ around the
  \ion{N}{5}~$\lambda$1239 line at \zabs = 2.9749.  The profiles of
  the \ion{Si}{4}~$\lambda$1394 line at \zabs\ = 2.8570 and the
  \ion{N}{5}~$\lambda$1243 line at \zabs\ = 2.9749 show a hint of time
  variability (see Figure~\ref{fig:velplot}).
  The weighted average velocity shift for class-A/B NALs ($\Delta
  v_{\rm AB}$ = 0.45~$\pm$~0.32~\kms) is not significantly shifted
  from that for class-C NALs ($\Delta v_{\rm C}$ =
  1.28~$\pm$~0.50~\kms).  Even if we consider only the reliable NALs
  (2 class-A/B lines and 2 class-C lines) that are classified as {\tt
    a1} in Table~\ref{tab:nal}, the difference would still be
  insignificant.  We conclude that there is no clear velocity shift of
  the class-A/B NALs between our observations.

\item[Q0450--1310] In the spectrum of this
  radio-quiet quasar, \citet{mis07a} detected 10 NALs (5 \ion{C}{4}
  and 5 \ion{Si}{4}), of which only the \ion{C}{4} NAL at \zabs\ =
  2.2307 (\vej\ $\sim$ 6370~\kms) is classified as intrinsic
  (class-A).  Because two \ion{C}{4} NALs at \zabs\ = 2.1050 and
  2.1061, and three \ion{Si}{4} NALs at \zabs\ = 2.1051, 2.1062, and
  2.1069 are close to each other within $\sim$170~\kms, we treat them
  as single NALs (\ion{C}{4} NAL at \zabs\ $\sim$ 2.1061 and
  \ion{Si}{4} NAL at \zabs\ $\sim$ 2.1059), respectively.  After
  removing NALs that are blended with data defects or other unrelated
  lines, we are left with 10 lines in 6 NALs.  The time interval
  between the observations is $\Delta t_{\rm obs}$ $\sim$ 17.1~yrs
  (i.e., $\Delta t_{\rm rest}$ $\sim$ 5.2~yrs).
  The weighted average velocity shift of the class-A NAL ($\Delta
  v_{\rm AB}$ = $-$3.30~$\pm$~0.38~\kms) is not shifted significantly
  from that of class-C NALs ($\Delta v_{\rm C}$ =
  $-$3.87~$\pm$~0.30~\kms).  Even after we consider only reliable NALs
  (2 class-A lines and 6 class-C lines) that satisfy all the criteria
  (see {\S}4), the difference is not significant.  We conclude that
  the class-A NAL is not significantly shifted between our
  observations.

\item[HE0940--1050] In total, 10 NALs (7 \ion{C}{4} and 3 \ion{Si}{4})
  were detected, of which only one \ion{Si}{4} NAL at \zabs\ = 2.8347
  (\vej\ $\sim$ 18600~\kms) is classified as class-B \citep{mis07a}.
  No class-A NALs were identified.  Although the \ion{C}{4} NAL at
  \zabs\ = 2.8346 was originally classified as class-C, it could be
  intrinsic to the quasar because i) it has a similar redshift to the
  class-B \ion{Si}{4} NAL and because ii) it shows line-locking with a
  \ion{C}{4} NAL at \zabs\ = 2.8245. We regard the two line-locked
  \ion{C}{4} NALs as a single NAL at \zabs\ $\sim$ 2.8294.  After
  removing NALs that are blended with data defects or other unrelated
  lines, we are left with 17 lines in 10 NALs.  The time interval
  between the observations is $\Delta t_{\rm obs}$ $\sim$ 15.8~yrs
  (i.e., $\Delta t_{\rm rest}$ $\sim$ 3.9~yrs).
  The weighted average velocity shift of class-A/B NALs ($\Delta
  v_{\rm AB}$ = $-$1.51~$\pm$~0.37~\kms) is not clearly shifted
  relative to that of class-C NALs ($\Delta v_{\rm C}$
  $-$0.94~$\pm$~0.12~\kms).  Even if we consider only reliable NALs (2
  class-B lines and 12 class-C lines), the difference would still be
  insignificant.  Thus, we conclude that the class-B NAL does not show
  a velocity shift between our observations compared to class-C NALs.

\item[HE1009$+$2956] Because of limited wavelength coverage,
  \citet{mis07a} detected only 3 NALs (1 \ion{C}{4}, 1 \ion{Si}{4},
  and 1 \ion{N}{5}), of which a \ion{Si}{4} NAL at \zabs\ = 2.2533
  (\vej\ $\sim$ 33900~\kms) and a \ion{N}{5} NAL at \zabs\ = 2.6495
  (\vej\ $\sim$ $-$450~\kms) are classified as class-A.
  Unfortunately, we cannot monitor the variability of the \ion{N}{5}
  NAL profile because of blending with other lines and a data defect
  in the first spectrum.  We used only three lines in two NALs for our
  analyses.  The time interval between the observations is $\Delta
  t_{\rm obs}$ $\sim$ 20.1~yrs (i.e., $\Delta t_{\rm rest}$ $\sim$
  5.5~yrs).
  The weighted average of the velocity shift of class-A/B NALs
  ($\Delta v_{\rm AB}$ = $-$1.67~$\pm$~1.07~\kms) is not significantly
  shifted compared to the class-C NALs ($\Delta v_{\rm C}$
  $-$0.68~$\pm$~0.49~\kms).  All detected NALs satisfy the criteria to
  be in our reliable sample.  Thus, we conclude that there is no clear
  velocity shift of the class-A \ion{Si}{4} NAL between our
  observations. For the quasar, we used equation~\ref{eqn:sigv_alt} to
  calculate the weighted average of the velocity shift for the class-A
  NAL because the sample includes only one member.

\item[Q1107$+$4847] \citet{mis07a} detected 6 NALs (3 \ion{C}{4} and 3
  \ion{Si}{4}), of which only one \ion{Si}{4} NAL at \zabs\ = 2.7243
  (\vej\ $\sim$ 21400~\kms) is classified as class-A.  There are no
  class-B NALs.  A \ion{C}{4} NAL at \zabs\ = 2.7243, originally
  classified as class-C, also could be intrinsic to the quasar because
  it has the same redshift as the class-A \ion{Si}{4} NAL.  Although
  all 6 NALs are covered by our spectra, four of the lines, including
  the \ion{C}{4} NAL at \zabs\ = 2.7243, are affected by data defects
  or are severely blended with other lines.  As a result, we used 8
  lines in 5 NALs for the CCF and the CCCD analyses.  The time
  interval between the observations is $\Delta t_{\rm obs}$ $\sim$
  20.7~yrs that corresponds to $\Delta t_{\rm rest}$ $\sim$ 5.2~yrs.
  The velocity shift of the class-A NAL ($\Delta v_{\rm AB}$ =
  $-$5.84~$\pm$~1.56~\kms) is not obviously shifted compared to the
  class-C NALs ($\Delta v_{\rm C}$ = $-$3.38~$\pm$~0.39~\kms).  The
  significance level becomes smaller when we remove NALs that do not
  satisfy all the selection criteria.
  Because all NALs for the quasar are blueshifted significantly (i.e.,
  $\Delta v$ $\leq$ $-$3~\kms), we speculate that the class-C NALs are
  also intrinsic to the quasar and accelerated with the class-A NAL.
  Indeed, some intrinsic NALs could have full coverage (i.e.,
  classified into class-C) if absorbing clouds in the vicinity of the
  continuum source are large enough to cover it completely
  \citep{mis07a}.  To test the idea, we performed the CCF and CCCD
  analyses to a \ion{Mg}{2}~$\lambda$2803 line at \zabs\ = 0.8076
  whose origin should be unrelated to the outflowing winds because its
  offset velocity would be $>$0.66c if we assume it is an intrinsic
  NAL.  We derived an offset velocity of $\Delta v$ =
  $-$5.61~$\pm$~0.28~\kms\ for the \ion{Mg}{2} NAL that is close to
  those of class-A/B and class-C NALs.
  Thus, we do not find any significant velocity shifts between our
  observations.  Here, we used equation~\ref{eqn:sigv_alt} to
  calculate the weighted average of the velocity shift for the class-A
  NAL because the sample includes only one member.

\item[HS1700$+$6416] This extremely bright quasar could be amplified
  by gravitational lensing by two clusters of galaxies along our
  sight-line. \citet{mis07a} detected 10 NALs (6 \ion{C}{4}, 2
  \ion{Si}{4}, and 2 \ion{N}{5}), of which three \ion{C}{4} NALs at
  \zabs\ = 2.4330, 2.4394, and 2.7125, and two \ion{N}{5} NALs at
  \zabs\ = 2.7125 and 2.7164 are classified as class-A or B.  A
  \ion{C}{4} NAL at \zabs\ = 2.1680 is not covered by our second-epoch
  spectrum. Because two \ion{C}{4} NALs at \zabs\ = 2.4330 and 2.4394
  show line-locking, we treat them as a single NAL at \zabs\ $\sim$
  2.4390. Thus, we used 11 lines in 8 NALs for the analyses. The time
  interval of the observations is $\Delta t_{\rm obs}$ $\sim$ 10.3~yrs
  (i.e., $\Delta t_{\rm rest}$ $\sim$ 2.8~yrs).
  The weighted average velocity shift of the class-A/B NALs ($\Delta
  v_{\rm AB}$ = 1.12~$\pm$~0.45~\kms) is not significantly shifted
  from that for class-C NALs ($\Delta v_{\rm C}$ =
  0.56~$\pm$~0.31~\kms).  Even if we consider only the reliable NALs
  (3 class-A lines and 5 class-C lines) that satisfy all the selection
  criteria, the significance level would still be insignificant.  We
  conclude that we do not detect a velocity shift between our
  observations.
\end{description}

The main result of the CCF and the CCCD analyses is that we do not
detect significant weighted average velocity shifts of class-A/B
relative to class-C NALs, placing a 3$\sigma$ upper limit on the
acceleration/deceleration of intrinsic NALs with a magnitude of
$\sim$0.7~km~s$^{-1}$~yr$^{-1}$ ($\sim$0.002~cm~s$^{-2}$) in average.

\section{Discussion}
We monitored intrinsic NALs in 6 quasars for 2.8--5.5~yr in the quasar
rest frame. We found no highly significant shifts in NAL velocities,
implying a 3$\sigma$ upper limits on the acceleration of
$\sim$0.7~km~s$^{-1}$~yr$^{-1}$ ($\sim$0.002~cm~s$^{-2}$).  The limits
on NAL acceleration found here appear to be two or three orders of
magnitude lower than those of BALs reported in the literature
\citep{jos18,gri16,jos14,hal07,gab03,rup02,vil01}. This is
particularly noteworthy since the NAL quasars in our sample have
larger bolometric luminosities than the BAL quasars reported in the
literature.

The simplest hypothesis for the acceleration of NAL systems invokes
radiation pressure. Thus, the equation of motion can be cast as
\begin{equation}
\frac{dv}{dt} = \frac{fL}{4\pi r^2 cm_p N_H} - \frac{GM}{r^2},
\label{eqn:dvdt}
\end{equation}
where $v$ is the outflow velocity, $L$ is the quasar's bolometric
luminosity, $f$ is the fraction of the luminosity that contributes to
the acceleration, $r$ is a distance from the flux source, $N_H$ is the
total column density of the absorber, and $M$ is the mass of the
central black-hole \citep{ham98}.  By considering the typical
luminosity and black hole mass of our target quasars (i.e., $L$ =
10$^{48}$~erg~s$^{-1}$, $M$ = 10$^{9}~M_{\odot}$) as well as a typical
hydrogen column density of NAL absorbers ($N_H$ = 10$^{18}$~\cmm;
e.g., \citealt{wu10}), equation~\ref{eqn:dvdt} becomes
\begin{equation}
\begin{split}
  \Delta v &\sim\ 52~f~\cos\theta \;
  \left(\frac{\Delta t}{\rm 1\; yr}\right)
  \left(\frac{r}{\rm 1\;kpc}\right)^{-2} \\
& \quad\times\left(\frac{L}{\rm 10^{48}\;erg\;s^{-1}}\right)
  \left(\frac{N_H}{\rm 10^{18}\;cm^{-2}}\right)^{-1}
    ~{\rm km~s}^{-1},
\label{eqn:dvdt2}
\end{split}
\end{equation}
where we ignore gravity because its contribution is much smaller than
the radiative force as long as $f > 10^{-7}$.  We also
include a factor of $\cos\theta$, where $\theta$ is the angle between the
outflow stream lines and our sightline.

\begin{figure*}
 \begin{center}
  \includegraphics[width=15cm,angle=0]{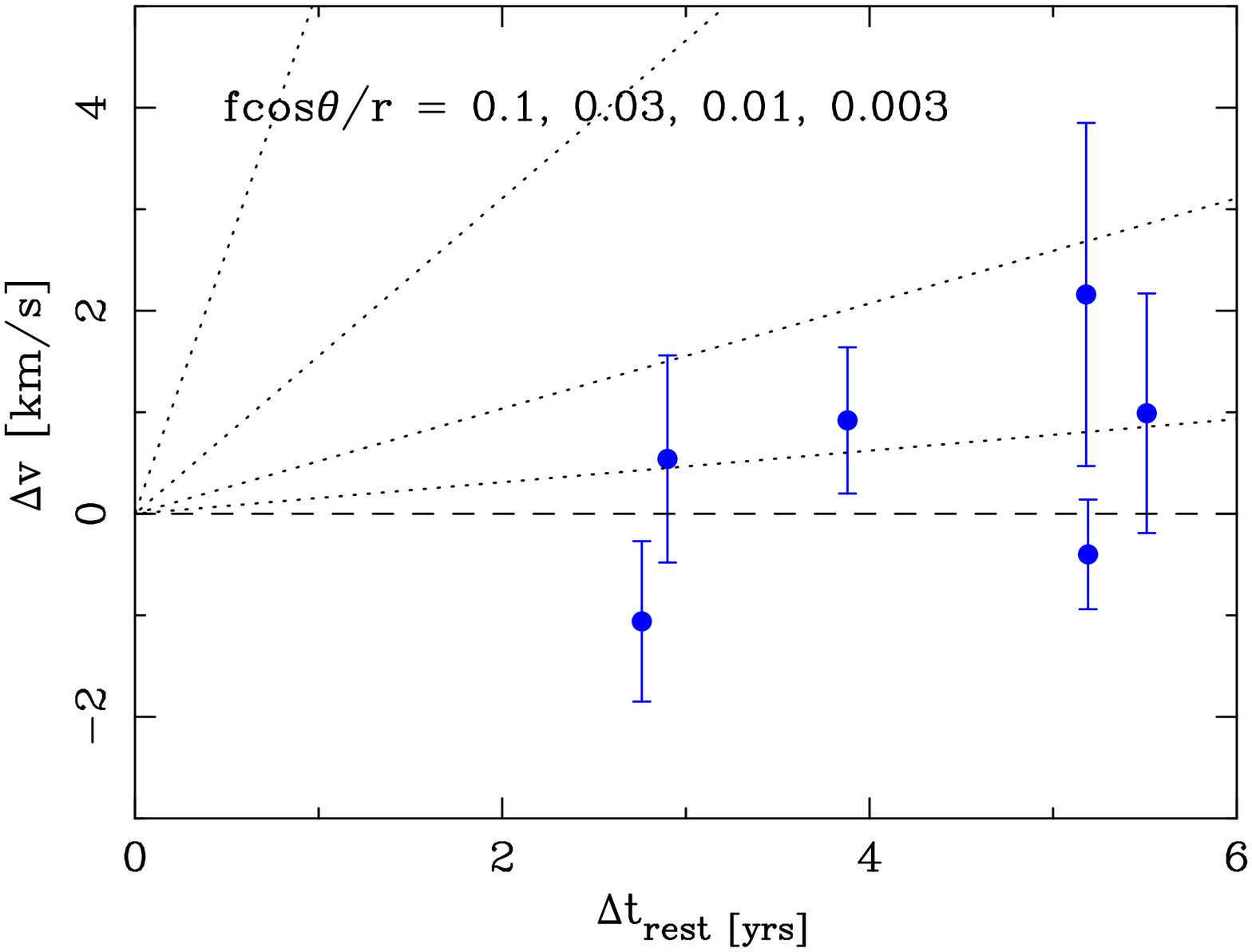}
 \end{center}
\caption{Velocity shift of intrinsic NALs between the observations as
  a function of time interval (filled blue circles).  The four dashed
  lines are the predicted velocity shifts from
  equation~\ref{eqn:dvdt2} with $f \cos\theta / r$ = 0.1, 0.03, 0.01,
  or 0.003 (from top to bottom), for the typical luminosity
  and black hole mass of our target quasars ($L$ =
  10$^{48}$~erg~s$^{-1}$, $M$ = 10$^{9}~M_{\odot}$) as well as a
  typical hydrogen column density of NAL absorbers ($N_H$ =
  10$^{18}$~\cmm).\label{fig:dv-dt}}
\end{figure*}

The implied acceleration of intrinsic NALs for our 6 quasars is
smaller than the prediction of equation~\ref{eqn:dvdt} by a factor of
100 or more, if we assume a distance of NAL absorbers from the central
engine of $r$ $\sim$ 1~kpc, following \citet{ara13}, for example.
Thus, if radiation pressure is the main acceleration mechanism, we
infer that, either $f ~\cos\theta \leq 10^{-2}$, or $r > 10$~kpc, or
both (see Figure~\ref{fig:dv-dt}). Because reports of absorber
distances as large as 10~kpc are uncommon in the literature, we
speculate that one or a combination of the following is the cause for
the small inferred accelerations: (i) $f < 10^{-2}$, which is much
smaller than the value suggested for X-ray Ultra Fast Outflows ($f$
$\geq$ 0.05; e.g., \citealt{tom13}), (ii) Our sightline toward the
flux source is almost perpendicular to some of the outflow stream
lines as suggested in \citet{ara99}, (iii) The acceleration is
episodic and it is only observable in a small fraction of quasars at
any given time. The last hypothesis is motivated by the results of
\citet{gri16} who were able to detect an acceleration in only three
out of 140 quasars and noted that the acceleration was not constant in
time. If this is the case for NALs as well, we would need to observe a
much larger sample of objects or make many repeated observations of
this sample in order to detect NAL acceleration.

\section{Summary}
In this study, we used high-resolution spectra of 6 NAL quasars that
were obtained with 8--10 meter telescopes (i.e., Subaru, Keck, and
VLT) at two epochs separated by $\Delta t_{\rm obs}$ $\sim$ 10.3--20.7
years (i.e., $\Delta t_{\rm rest}$ $\sim$ 2.8--5.5 years) to monitor
velocity changes of {\it intrinsic} NALs.  Our main results are as
follows.

\begin{itemize}
\item Using the CCF and CCCD analyses, we discovered no significant
  acceleration/deceleration of intrinsic NALs, placing a 3$\sigma$
  upper limit on the magnitude of the acceleration of
  $\sim$0.7~km~s$^{-1}$~yr$^{-1}$ ($\sim$0.002~cm~s$^{-2}$).
\item This magnitude of NAL acceleration is more than two orders of
  magnitude smaller than the expected value from radiative
  acceleration, which suggests that, if our sightline is almost
  parallel to the streamlines of the outflow, the fraction of the
  luminosity that contributes to the acceleration is $f < 0.01$, or
  the absorber's distance from the flux source is greater than
  10~kpc. We do not find any correlations between gas acceleration and
  bolometric luminosity either in BAL or in NAL quasars.
\item Because of the very small velocity shifts, the NAL absorbers
  studied here may be at or close to their {\it terminal} velocities,
  i.e., they may be located at a large distance from the quasar
  central engine.  Other possible explanations are that we observe a
  standing-flow with streamlines that are almost perpendicular to our
  sightline and gas is continuously replenished as suggested by
  hydrodynamical simulations \citep{ara99,pro00}, or that the
  acceleration process is episodic so that only a small fraction of
  NALs exhibit acceleration at any one time.
\end{itemize}

If the typical magnitude of NAL acceleration is close to the 3$\sigma$
upper limit we have obtained (i.e., $\sim$0.7~km~s$^{-1}$~yr$^{-1}$),
we may be able to detect obvious velocity shifts in future
observations compared to our first epoch spectra taken in 1995--2000.
For example, intrinsic NALs are expected to be blueshifted by more
than 2~\kms\ with a significance level of $\geq$ 3$\sigma$ in 40 years
from now. Alternatively, if the acceleration process is episodic, we
may be able to detect acceleration in a few quasars if we observe a
sample that is an order of magnitude larger than the sample we have
used here.  Such a precise spectroscopic observation will also be very
important to other areas of research, such as a direct measurement of
cosmic expansion \citep[e.g.,][]{san62,loe98,bal07,lis08} and a
measurement of changes in the fine structure constant
\citep[e.g.,][]{mur03,cha04,sri04,web11} using extremely large (30-m
class) telescopes in coming decades.

\acknowledgments The research was supported by the Japan Society for
the Promotion of Science through Grants-in-Aid for Scientific Research
18K03698 and partially supported by a MEXT Grant-in-Aid for Scientific
Research on Innovative Areas (No.15H05894).


\clearpage
\begin{turnpage}
\begin{deluxetable*}{ccccccccccccc}
\tabletypesize{\scriptsize}
\tablecaption{Sample Quasars\label{tab:qso}}
\tablewidth{0pt}
\tablehead{
\colhead{Quasar}         &
\colhead{RA}             &
\colhead{Dec}            &
\colhead{\zem}           &
\colhead{$m_{\rm V}$}      &
\colhead{$L_{\rm bol}$$^a$} &
\colhead{$R^b$}           &
\colhead{Obs Date}       &
\colhead{Tel./Inst.}     &
\colhead{$R$}            &
\colhead{$T_{\rm exp}$}    &
\colhead{S/N$^c$}        &
\colhead{$\Delta t_{\rm rest}$$^d$}  \\
\colhead{}               &
\colhead{(hh mm ss)}     &
\colhead{(dd mm ss)}     &
\colhead{}               &
\colhead{(mag)}          &
\colhead{(ergs s$^{-1}$)} &
\colhead{}               &
\colhead{}               &
\colhead{}               &
\colhead{}               &
\colhead{(sec)}          &
\colhead{(pixel$^{-1}$)}  &
\colhead{(yrs)}          
}
\startdata
HE0130--4021 & 01 33 02 & -40 06 28 & 3.030 & 17.02 & 1.1$\times$10$^{48}$ & 11.2   & 1995 Dec. 28    & Keck/HIRES & 36000 & 14400 & 41 & 2.9 \\
            &          &           &       &       &                     &        & 2007 Sep.  5    & VLT/UVES   & 40000 &  6000 & 46 &     \\
 Q0450--1310 & 04 53 12 & -13 05 46 & 2.300 & 16.50 & 1.1$\times$10$^{48}$ & $<$1.7 & 1998 Dec. 13-14 & Keck/HIRES & 36000 &  7200 & 19 & 5.2 \\
            &          &           &       &       &                     &        & 2016 Jan. 27    & Subaru/HDS & 45000 & 10800 & 15 &     \\
HE0940--1050 & 09 42 53 & -11 04 25 & 3.080 & 16.90 & 1.3$\times$10$^{48}$ & $<$2.6 & 2000 Apr.  3    & VLT/UVES   & 40000 &  3600 & 32 & 3.9 \\
            &          &           &       &       &                     &        & 2016 Jan. 27    & Subaru/HDS & 45000 &  7200 & 29 &     \\
HE1009+2956 & 10 11 55 & +29 41 41 & 2.644 & 16.40 & 1.5$\times$10$^{48}$ & $<$1.6 & 1995 Dec. 29    & Keck/HIRES & 36000 & 12200 & 56 & 5.5 \\
            &          &           &       &       &                     &        & 2016 Jan. 27    & Subaru/HDS & 45000 &  7200 & 27 &     \\
 Q1107+4847 & 11 10 38 & +48 31 16 & 3.000 & 16.60 & 1.6$\times$10$^{48}$ & $<$2.0 & 1995 May   9    & Keck/HIRES & 37500 &  7200 & 59 & 5.2 \\
            &          &           &       &       &                     &        & 2016 Jan. 27    & Subaru/HDS & 45000 &  7200 & 37 &     \\
HS1700+6416 & 17 01 00 & +64 12 09 & 2.722 & 16.13 & 2.1$\times$10$^{48}$ & $<$1.2 & 1995 May  10    & Keck/HIRES & 36000 & 11500 & 72 & 2.8 \\
            &          &           &       &       &                     &        & 2005 Aug. 19    & Subaru/HDS & 36000 &  5400 & 25 &     \\
\enddata
\tablenotetext{a}{Bolometric luminosity, calculated as $L_{\rm bol}$ = 4.4$\lambda L_{\lambda}$(1450\AA), following \citet{ric06}.}
\tablenotetext{b}{Radio-loudness, calculated as $R$ = $f_{\nu}$(5 GHz)/$f_{\nu}$(4400\AA), following \citet{kel89,kel94}.}
\tablenotetext{c}{Signal to noise ratio at $\lambda_{\rm obs}$ $\sim$4900\AA.}
\tablenotetext{d}{Time interval between observations in the quasar's rest-frame.}
\end{deluxetable*}
\end{turnpage}

\clearpage
\begin{landscape}
\LongTables
\begin{deluxetable*}{ccccccccccl}
\tablecaption{Sample NALs\label{tab:nal}}
\tablewidth{0pt}
\tablehead{
\colhead{Quasar} & 
\colhead{\zabs} &
\colhead{\vej$^a$} &
\colhead{ion} &
\colhead{comp.$^b$} &
\colhead{class$^c$} &
\colhead{$\Delta v$(CCF)$^d$} &
\colhead{$\Delta v$(CCCD)$^e$} &
\colhead{$\Delta v \pm \sigma(\Delta v)^f$} &
\colhead{sample$^g$} &
\colhead{notes} \\
\colhead{} &
\colhead{} &
\colhead{(\kms)} &
\colhead{} &
\colhead{} &
\colhead{} &
\colhead{(\kms)} &
\colhead{(\kms)} &
\colhead{(\kms)} &
\colhead{} &
\colhead{} 
}
\startdata
HE0130--4021 & 2.2316 & 65181 & \ion{C}{4}  & b & B1 &       &                &                & d  & blend with other lines. \\
            &        &       &             & r &    &  1.41 &  1.34$\pm$0.34 &  1.34$\pm$1.30 & a1 & \\
            & 2.5597 & 37037 & \ion{C}{4}  & b & A2 &  0.29 &  0.26$\pm$0.21 &  0.26$\pm$1.27 & a1 & \\
            &        &       &             & r &    &       &                &                & d  & affected by data defect in 1st spectrum. \\
            & 2.6884 & 26503 & \ion{C}{4}  & b & C2 &       &                &                & d  & affected by data defect in 1st spectrum. \\
            &        &       &             & r &    &  2.27 &  2.24$\pm$0.58 &  2.24$\pm$1.38 & a1 & \\
            & 2.8570 & 13155 & \ion{Si}{4} & b & C2 &  1.19 &  1.19$\pm$0.30 &  1.19$\pm$1.29 & c  & may be time variable. \\
            &        &       &             & r &    &       &                &                & d  & blend with other lines. \\
            & 2.8571 & 13147 & \ion{C}{4}  & b & C2 &  0.38 &  0.49$\pm$0.44 &  0.49$\pm$1.33 & a1 & \\
            &        &       &             & r &    &       &                &                & d  & affected by data defect in 1st spectrum. \\
            & 2.9749 &  4130 & \ion{N}{5}  & b & A2 & -0.16 & -0.14$\pm$0.35 & -0.14$\pm$1.30 & b  & margin is $<$2\AA\ because of data defect. \\
            &        &       &             & r &    &  0.40 &  0.34$\pm$0.49 &  0.34$\pm$1.34 & c  & may be time variable. \\
\hline                   
Q0450--1310  & 1.6968 & 59748 & \ion{C}{4}  & b & C2 & -4.20 & -4.26$\pm$1.40 & -4.26$\pm$2.18 & b  & margin is $<$ 2\AA\ because of line blending. \\
            &        &       &             & r &    & -1.05 & -0.91$\pm$2.53 & -0.91$\pm$3.03 & a1 & \\
            & 1.9986 & 28646 & \ion{C}{4}  & b & C2 & -3.44 & -3.47$\pm$0.70 & -3.47$\pm$1.81 & a1 & \\
            &        &       &             & r &    & -4.66 & -4.63$\pm$1.41 & -4.63$\pm$2.19 & a1 & \\
            & 2.0667 & 21957 & \ion{Si}{4} & b & C1 & -4.10 & -4.21$\pm$0.83 & -4.21$\pm$1.86 & a1 & \\
            &        &       &             & r &    &       &                &                & d  & affected by data defect in 1st spectrum. \\
            & 2.0668 & 21947 & \ion{C}{4}  & b & C1 & -3.67 & -3.76$\pm$0.81 & -3.76$\pm$1.86 & a1 & \\
            &        &       &             & r &    &       &                &                & d  & affected by data defect in 1st and 4th spectra. \\
            & 2.1059 & 18163 & \ion{Si}{4} & b & C2 &       &                &                & d  & blend with other lines. \\
            &        &       &             & r &    &       &                &                & d  & affected by data defect in 1st spectrum. \\
            & 2.1061 & 18144 & \ion{C}{4}  & b & C1 & -3.81 & -3.75$\pm$0.39 & -3.75$\pm$1.71 & a1 & \\
            &        &       &             & r &    & -4.43 & -4.41$\pm$0.42 & -4.41$\pm$1.72 & b  & margin is $<$2\AA\ because of data defect. \\
            & 2.2307 &  6366 & \ion{Si}{4} & b & A2 & -3.51 & -3.65$\pm$0.87 & -3.65$\pm$1.88 & a1 & \\
            &        &       &             & r &    & -3.14 & -2.89$\pm$1.12 & -2.89$\pm$2.01 & a1 & \\
\hline                   
HE0940--1050 & 2.2210 & 69629 & \ion{C}{4}  & b & C1 & -0.78 & -0.81$\pm$0.18 & -0.81$\pm$0.92 & a1 & \\
            &        &       &             & r &    & -1.16 & -1.15$\pm$0.24 & -1.15$\pm$0.93 & a1 & \\
            & 2.3302 & 60096 & \ion{C}{4}  & b & C1 & -0.91 & -0.89$\pm$0.19 & -0.89$\pm$0.92 & a1 & \\
            &        &       &             & r &    & -1.13 & -1.14$\pm$0.18 & -1.14$\pm$0.92 & a1 & \\
            & 2.4090 & 53331 & \ion{C}{4}  & b & C3 & -0.95 & -0.89$\pm$0.41 & -0.89$\pm$0.99 & a1 & \\
            &        &       &             & r &    & -1.60 & -1.62$\pm$0.73 & -1.62$\pm$1.16 & a1 & \\
            & 2.6580 & 32625 & \ion{C}{4}  & b & C2 & -0.95 & -0.95$\pm$0.71 & -0.95$\pm$1.15 & a1 & \\
            &        &       &             & r &    & -0.59 & -0.48$\pm$0.54 & -0.48$\pm$1.05 & a1 & \\
            & 2.6677 & 31839 & \ion{C}{4}  & b & C1 & -1.07 & -1.01$\pm$0.57 & -1.01$\pm$1.07 & a1 & \\
            &        &       &             & r &    &  0.07 &  0.13$\pm$1.00 &  0.13$\pm$1.35 & a1 & \\
            & 2.6677 & 31839 & \ion{Si}{4} & b & C2 &       &                &                & d  & affected by data defect in 1st spectrum. \\
            &        &       &             & r &    & -2.72 & -2.60$\pm$2.35 & -2.60$\pm$2.52 & a1 & \\
            & 2.8245 & 19374 & \ion{C}{4}  & b & C2 & -0.31 & -0.27$\pm$0.94 & -0.27$\pm$1.30 & b  & margin is $<$2\AA\ because of line blending. \\
            &        &       &             & r &    & -1.06 & -0.98$\pm$2.03 & -0.98$\pm$2.22 & b  & margin is $<$2\AA\ because of line blending. \\
            & 2.8283 & 19077 & \ion{Si}{4} & b & C2 & -1.37 & -1.32$\pm$0.52 & -1.32$\pm$1.04 & a1 & \\
            &        &       &             & r &    &       &                &                & d  & affected by data defect in 1st spectrum. \\
            & 2.8294 & 18991 & \ion{C}{4}  &b,r& B2 & -0.82 & -0.83$\pm$0.23 & -0.83$\pm$0.93 & a2 & consider blue and red components together because of line locking. \\
            & 2.8347 & 18578 & \ion{Si}{4} & b & B1 & -1.95 & -1.92$\pm$0.35 & -1.92$\pm$0.97 & a1 & \\
            &        &       &             & r &    & -1.85 & -1.88$\pm$0.53 & -1.88$\pm$1.04 & a1 & \\
\hline                   
HE1009+2956 & 1.9065 & 66707 & \ion{C}{4}  & b & C1 & -1.11 & -1.13$\pm$0.37 & -1.13$\pm$1.06 & a1 & \\
            &        &       &             & r &    & -0.22 & -0.14$\pm$0.59 & -0.14$\pm$1.15 & a1 & \\
            & 2.2533 & 33879 & \ion{Si}{4} & b & A2 &       &                &                & d  & blend with other lines. \\
            &        &       &             & r &    & -1.71 & -1.67$\pm$0.40 & -1.67$\pm$1.07 & a1 & \\
            & 2.6495 &$-$452 & \ion{N}{5}  & b & A2 &       &                &                & d  & affected by data defect in 1st spectrum. \\
            &        &       &             & r &    &       &                &                & d  & blend with other lines. \\
\hline                   
Q1107+4847  & 2.1433 & 70938 & \ion{C}{4}  & b & C2 & -5.01 & -5.09$\pm$0.20 & -5.09$\pm$1.46 & a1 & \\
            &        &       &             & r &    & -4.47 & -4.44$\pm$0.42 & -4.44$\pm$1.51 & a1 & \\
            & 2.7243 & 21388 & \ion{C}{4}  & b & C1 &       &                &                & d  & affected by data defect in 2nd spectrum. \\
            &        &       &             & r &    &       &                &                & d  & affected by data defect in 2nd spectrum. \\
            & 2.7243 & 21388 & \ion{Si}{4} & b & A2 & -5.78 & -5.84$\pm$0.58 & -5.84$\pm$1.56 & a1 & \\
            &        &       &             & r &    &       &                &                & d  & blend with other lines. \\
            & 2.7593 & 18595 & \ion{Si}{4} & b & C1 & -3.20 & -3.21$\pm$0.29 & -3.21$\pm$1.48 & b  & margin is $<$2\AA\ because of line blending. \\
            &        &       &             & r &    & -2.61 & -2.52$\pm$0.39 & -2.52$\pm$1.50 & b  & margin is $<$2\AA\ because of line blending. \\
            & 2.7610 & 18460 & \ion{C}{4}  & b & C1 & -2.99 & -2.99$\pm$0.15 & -2.99$\pm$1.46 & a1 & \\
            &        &       &             & r &    & -2.24 & -2.25$\pm$0.13 & -2.25$\pm$1.46 & a1 & \\
            & 2.7621 & 18372 & \ion{Si}{4} & b & C2 & -3.15 & -3.16$\pm$0.22 & -3.16$\pm$1.47 & b  & margin is $<$ 2\AA\ because of line blending. \\
            &        &       &             & r &    &       &                &                & d  & affected by data defect in 1st spectrum. \\
\hline                   
HS1700+6416 & 2.3154 & 34551 & \ion{Si}{4} & b & C1 &  0.87 &  0.87$\pm$0.14 &  0.87$\pm$1.19 & a1 & \\
            &        &       &             & r &    &  0.25 &  0.25$\pm$0.15 &  0.25$\pm$1.19 & a1 & \\
            & 2.3156 & 34533 & \ion{C}{4}  & b & C1 &       &                &                & d  & affected by data defect in 1st spectrum. \\
            &        &       &             & r &    & -0.40 & -0.39$\pm$0.16 & -0.39$\pm$1.19 & a1 & \\
            & 2.4333 & 24169 & \ion{Si}{4} & b & C2 &  0.39 &  0.79$\pm$2.21 &  0.79$\pm$2.51 & c  & may be time variable. \\
            &        &       &             & r &    &       &                &                & d  & blend with other lines. \\
            & 2.4390 & 23675 & \ion{C}{4}  &b,r& B2 &  0.82 &  0.85$\pm$0.26 &  0.85$\pm$1.21 & a2 & consider blue and red components together because of line locking. \\
            & 2.5785 & 11789 & \ion{C}{4}  & b & C2 &  1.22 &  1.13$\pm$0.64 &  1.13$\pm$1.34 & a1 & \\
            &        &       &             & r &    &  2.16 &  2.08$\pm$1.78 &  2.08$\pm$2.14 & a1 & \\
            & 2.7125 &   767 & \ion{C}{4}  & b & A2 &  0.31 &  0.32$\pm$0.40 &  0.32$\pm$1.25 & c  & may be time variable. \\
            &        &       &             & r &    &       &                &                & d  & affected by data defect in 2nd spectrum. \\
            & 2.7125 &   767 & \ion{N}{5}  & b & A2 &  2.64 &  2.65$\pm$0.47 &  2.65$\pm$1.27 & a1 & \\
            &        &       &             & r &    &  1.14 &  1.08$\pm$0.47 &  1.08$\pm$1.27 & a1 & \\
            & 2.7164 &   452 & \ion{N}{5}  & b & A2 & -0.21 & -0.20$\pm$2.08 & -0.20$\pm$2.39 & a1 & \\
            &        &       &             & r &    &       &                &                & d  & blend with other lines. \\
\enddata
\tablenotetext{a}{Offset velocity that is defined as positive if NALs
  are blueshifted from the quasar.}
\tablenotetext{b}{Blue~(b) or red~(r) component of doublet.}
\tablenotetext{c}{NAL class: Class~A, B, or C denote reliable,
  possible, or intervening NALs, respectively \citep{mis07a}.}
\tablenotetext{d}{Cross-Correlation Function \citep{gri16}.}
\tablenotetext{e}{Cross-Correlation Centroid Distribution
  \citep{gri16}.}
\tablenotetext{f}{Velocity shift and its uncertainty after adding an
  error from spectral distortion. See discussion in
  \S~\ref{sec:crosscorr} of the text.}
\tablenotetext{g}{Sample class: a1) satisfy all criteria, a2)
  line-locking, b) sampling spectral region is smaller than 2\AA\ at
  either or both side of NAL because of other lines or data defect, c)
  show a hint of time variability, and d) NAL itself blends other line
  or data defect.}
\end{deluxetable*}
\clearpage
\end{landscape}

\begin{deluxetable*}{ccc}
\tablecaption{Velocity Uncertainty due to Spectrum Distortion\label{tab:rms}}
\tablewidth{0pt}
\tablehead{
\colhead{Quasar}                   &
\colhead{Number of Pairs$^a$}      &
\colhead{$\sigma(\Delta v)_{\rm rms}$}  \\
\colhead{}                         &
\colhead{}                         &
\colhead{(km s$^{-1}$)}             
}
\startdata
HE0130--4021 &  3 & 1.25 \\
 Q0450--1310 & 28 & 1.67 \\
HE0940--1050 & 91 & 0.90 \\
HE1009+2956 &  1 & 0.99 \\
 Q1107+4847 & 21 & 1.45 \\
HS1700+6416 & 15 & 1.18 \\
\enddata
\tablenotetext{a}{Roor-mean-square of $|\Delta v|$ between all pairs
  of class-C lines.}
\end{deluxetable*}

\begin{deluxetable*}{cccccccccc}
\tablecaption{Shift Velocity of NALs\label{tab:wa}}
\tablewidth{0pt}
\tablehead{
\multicolumn{1}{c}{} &
\multicolumn{3}{c}{all NALs} &
\multicolumn{1}{c}{} &
\multicolumn{5}{c}{reliable NALs} \\
\cline{2-4} 
\cline{6-10}
\colhead{Quasar} &
\colhead{$\Delta v_{\rm AB}$$^a$} &
\colhead{$\Delta v_{\rm C}$$^b$} &
\colhead{s.l.$^c$} &
\colhead{} &
\colhead{$\Delta v_{\rm AB}$$^a$} &
\colhead{$\Delta v_{\rm C}$$^b$} &
\colhead{s.l.$^c$} &
\colhead{$\Delta v_{\rm AB} - \Delta v_{\rm C}$} &
\colhead{($\Delta v_{\rm AB} - \Delta v_{\rm C}$)/$\Delta t_{\rm rest}$} \\
\colhead{}                            &
\colhead{(\kms)}                      &
\colhead{(\kms)}                      &
\colhead{}                            &
\colhead{}                            &
\colhead{(\kms)}                      &
\colhead{(\kms)}                      &
\colhead{}                            &
\colhead{(\kms)}                      &
\colhead{(cm~s$^{-2}$)}                      
}
\startdata
HE0130--4021 &  0.45$\pm$0.32     &  1.28$\pm$0.50 & 1.40$\sigma$ & &  0.79$\pm$0.54    &  1.33$\pm$0.87 & 0.53$\sigma$ & $-$0.54~$\pm$~1.02 & $-$0.0006~$\pm$~0.0011 \\
 Q0450--1310 & -3.30$\pm$0.38     & -3.87$\pm$0.30 & 1.18$\sigma$ & & -3.30$\pm$0.38    & -3.70$\pm$0.38 & 0.74$\sigma$ & $+$0.40~$\pm$~0.54 & $+$0.0002~$\pm$~0.0003 \\
HE0940--1050 & -1.51$\pm$0.37     & -0.94$\pm$0.12 & 1.47$\sigma$ & & -1.90$\pm$0.71    & -0.98$\pm$0.13 & 1.27$\sigma$ & $-$0.92~$\pm$~0.72 & $-$0.0007~$\pm$~0.0006 \\
HE1009+2956 & -1.67$\pm$1.07$^d$ & -0.68$\pm$0.49 & 0.84$\sigma$ & & -1.67$\pm$1.07$^d$ & -0.68$\pm$0.49 & 0.84$\sigma$ & $-$0.99~$\pm$~1.18 & $-$0.0006~$\pm$~0.0007 \\
 Q1107+4847 & -5.84$\pm$1.56$^d$ & -3.38$\pm$0.39 & 1.53$\sigma$ & & -5.84$\pm$1.56$^d$ & -3.68$\pm$0.65 & 1.28$\sigma$ & $-$2.16~$\pm$~1.69 & $-$0.0013~$\pm$~0.0010 \\
HS1700+6416 &  1.12$\pm$0.45     &  0.56$\pm$0.31 & 1.02$\sigma$ & &  1.61$\pm$0.71    &  0.55$\pm$0.35 & 1.34$\sigma$ & $+$1.06~$\pm$~0.79 & $+$0.0012~$\pm$~0.0009 \\
\enddata
\tablenotetext{a}{Weighted average of shift velocity for class-A/B NALs.}
\tablenotetext{b}{Weighted average of shift velocity for class-C NALs.}
\tablenotetext{c}{Significance level of difference in velocity shift between class-A/B and C NALs.}
\tablenotetext{d}{This is a shift velocity and its uncertainty that is
  evaluated using equations~\ref{eqn:delv} and \ref{eqn:sigv_alt}
  because only one class-A/B NAL was detected in the quasar spectrum.}
\end{deluxetable*}


\begin{figure*}
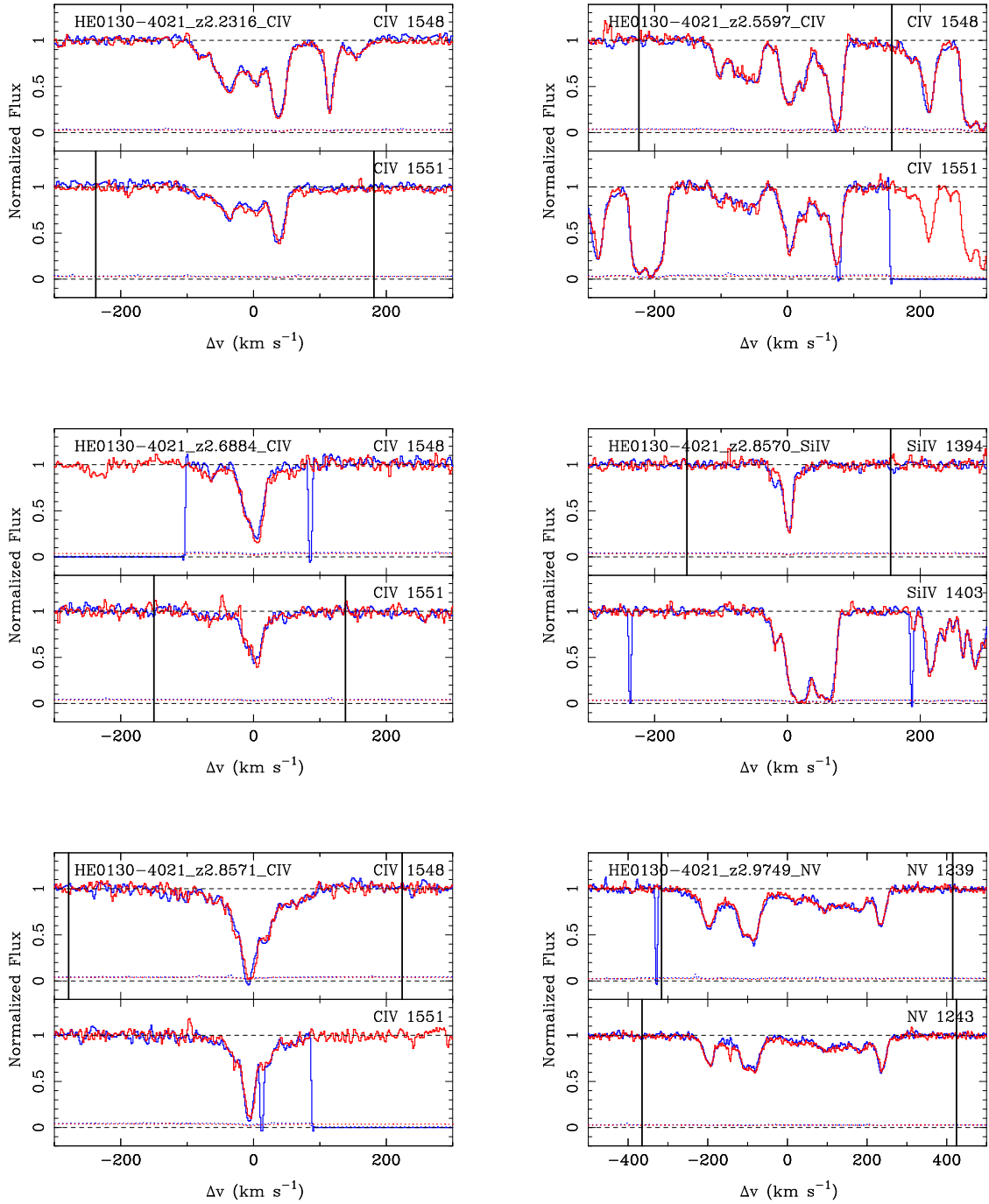

 \begin{center}
  \includegraphics[width=8cm,angle=0]{fig1_1.ps}
  \includegraphics[width=8cm,angle=0]{fig1_2.ps}
  \includegraphics[width=8cm,angle=0]{fig1_3.ps}
  \includegraphics[width=8cm,angle=0]{fig1_4.ps}
  \includegraphics[width=8cm,angle=0]{fig1_5.ps}
  \includegraphics[width=8cm,angle=0]{fig1_6.ps}
 \end{center}
\caption{Velocity plots of \ion{C}{4}, \ion{Si}{4}, and \ion{N}{5}
  NALs that are included in this work.  Blue and red histograms denote
  the observed spectra in the first and last epochs, respectively.
  Vertical lines mark the wavelength window over which we performed
  CCF and CCCD analyses.\label{fig:velplot}}
\end{figure*}

\begin{figure*}
\addtocounter{figure}{-1}
 \begin{center}
  \includegraphics[width=8cm,angle=0]{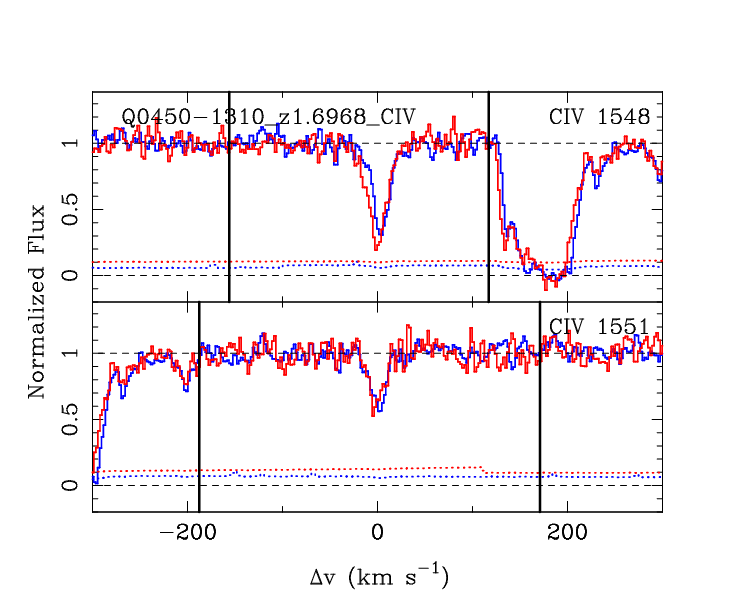}
  \includegraphics[width=8cm,angle=0]{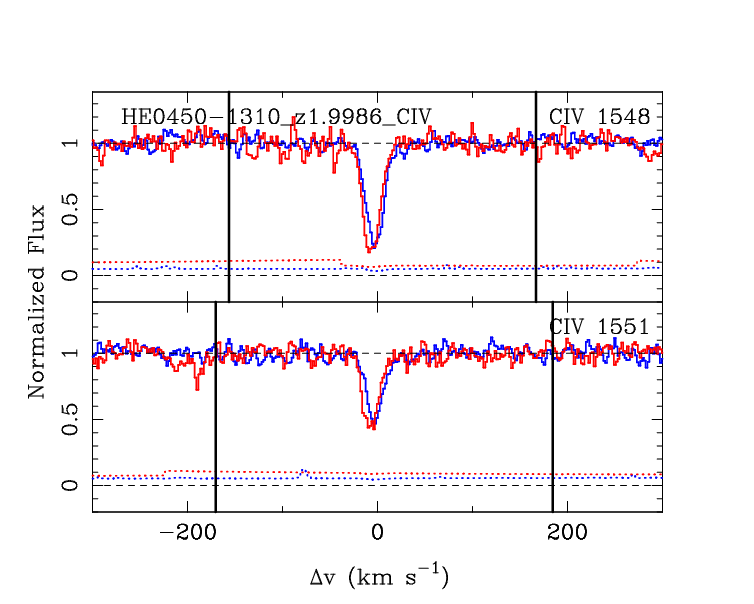}
  \includegraphics[width=8cm,angle=0]{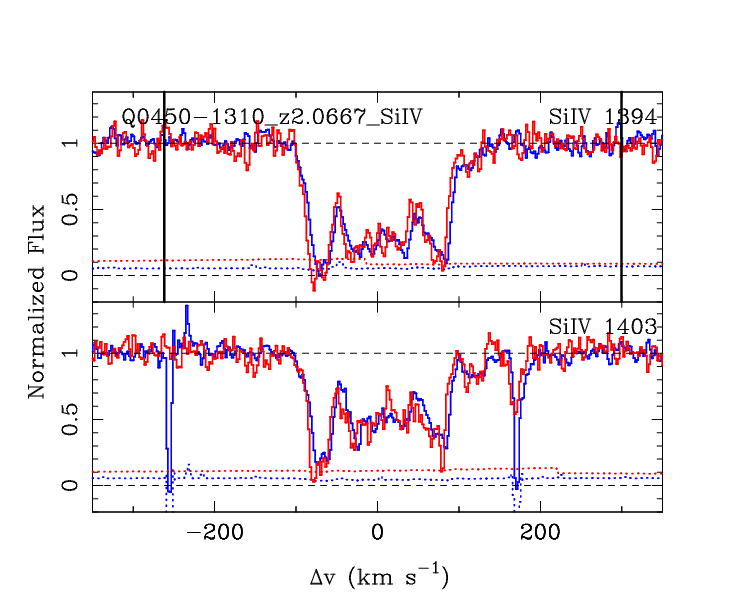}
  \includegraphics[width=8cm,angle=0]{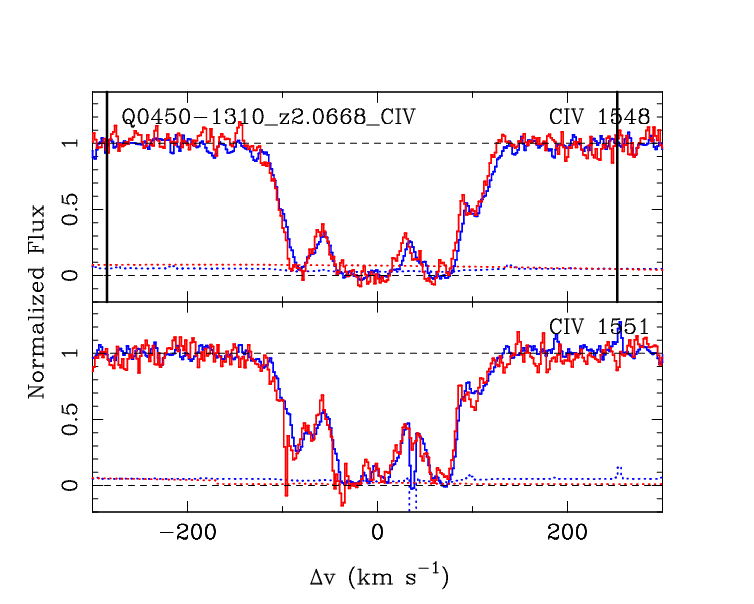}
  \includegraphics[width=8cm,angle=0]{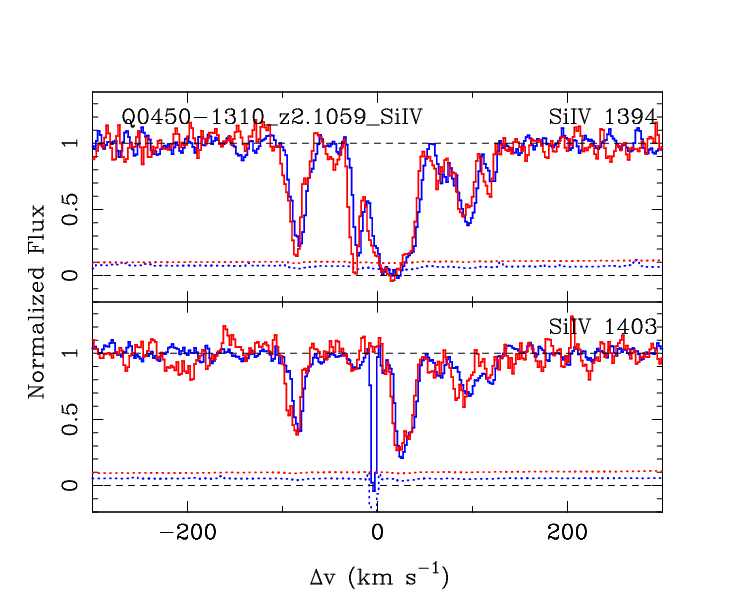}
  \includegraphics[width=8cm,angle=0]{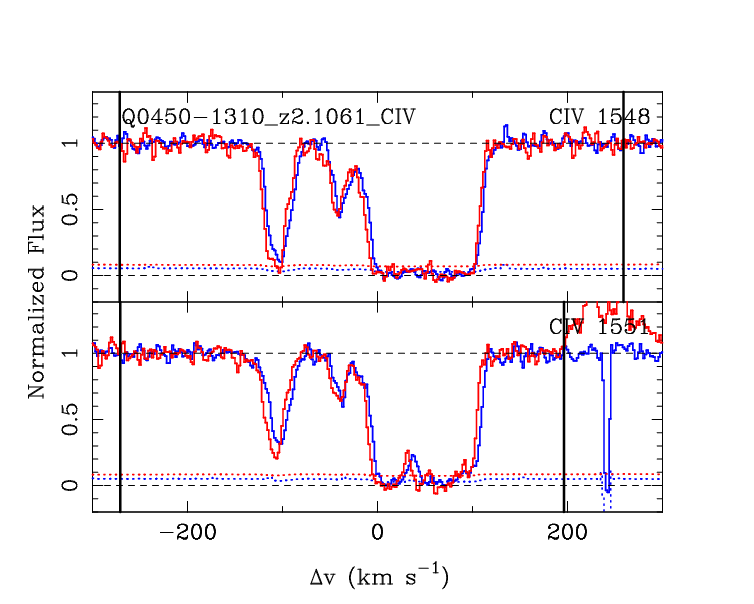}
 \end{center}
\caption{continued.}
\end{figure*}

\begin{figure*}
\addtocounter{figure}{-1}
 \begin{center}
  \includegraphics[width=8cm,angle=0]{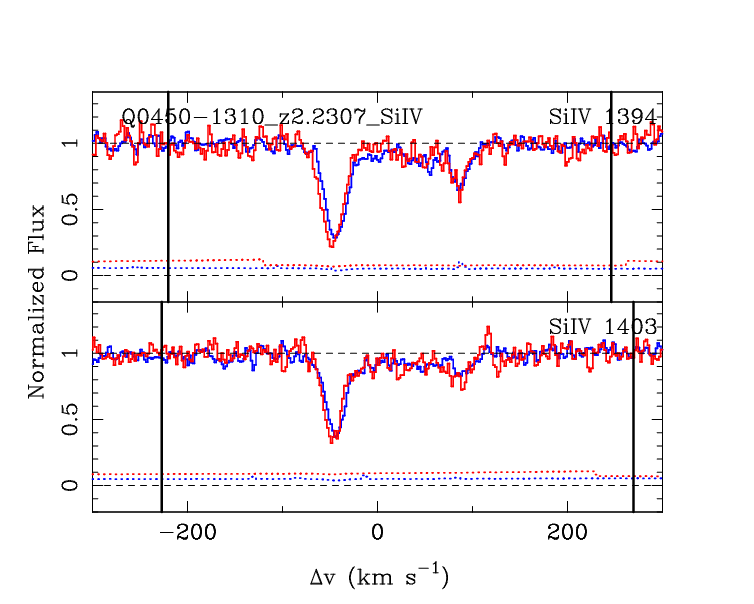}
  \includegraphics[width=8cm,angle=0]{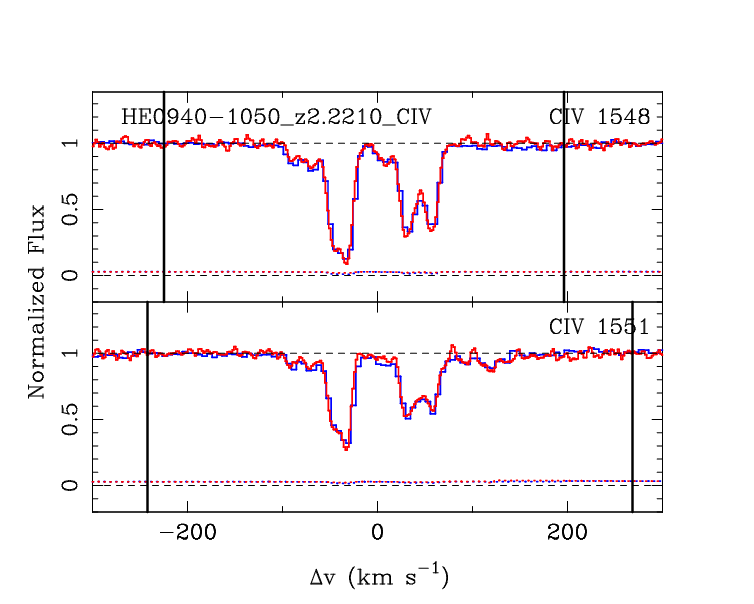}
  \includegraphics[width=8cm,angle=0]{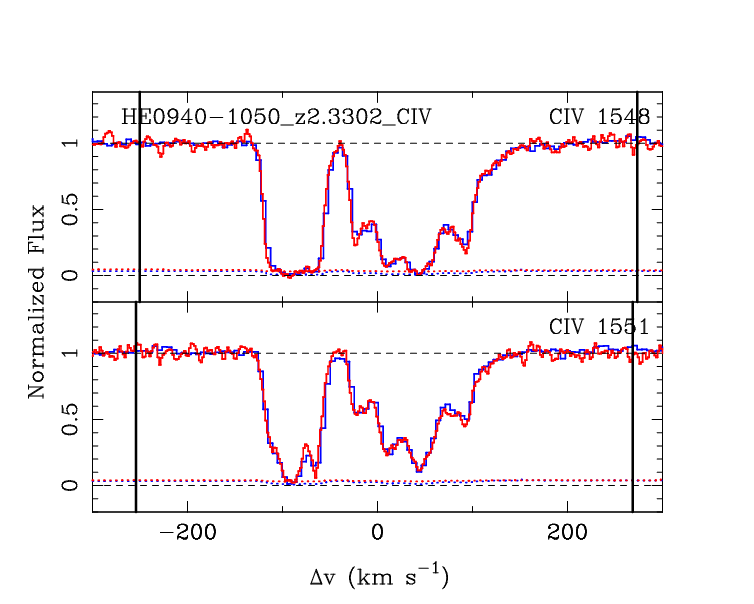}
  \includegraphics[width=8cm,angle=0]{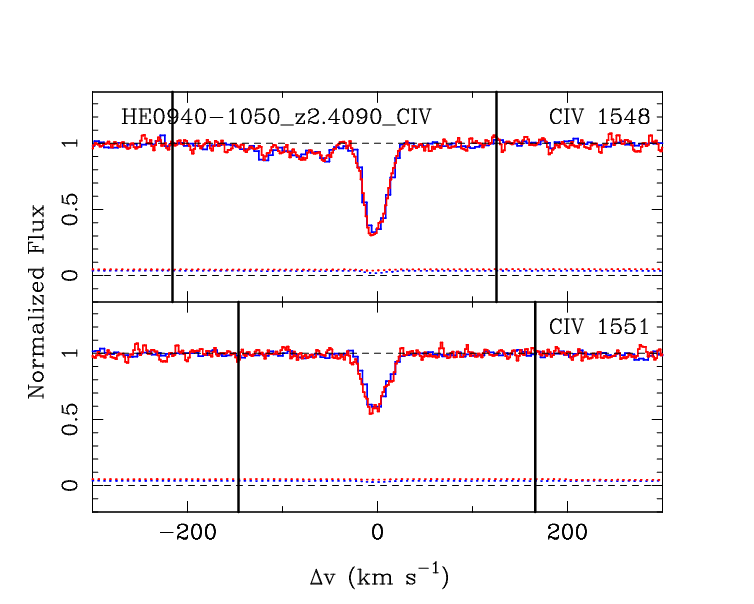}
  \includegraphics[width=8cm,angle=0]{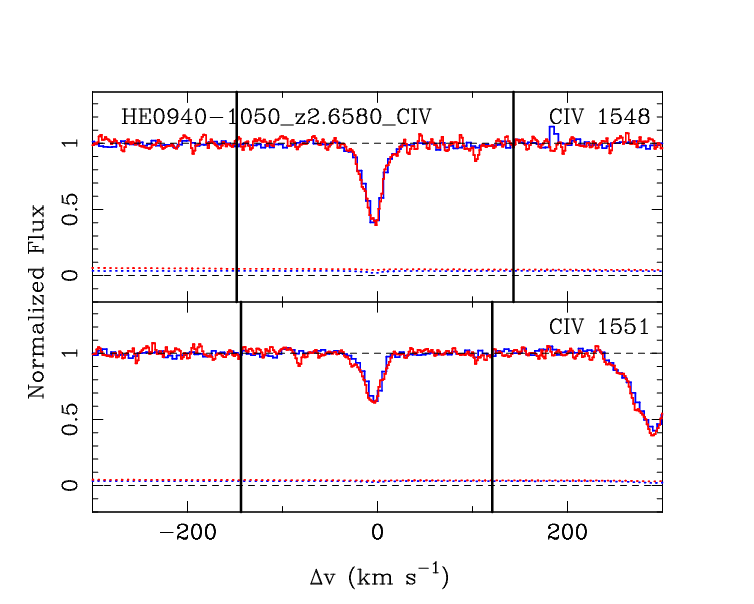}
  \includegraphics[width=8cm,angle=0]{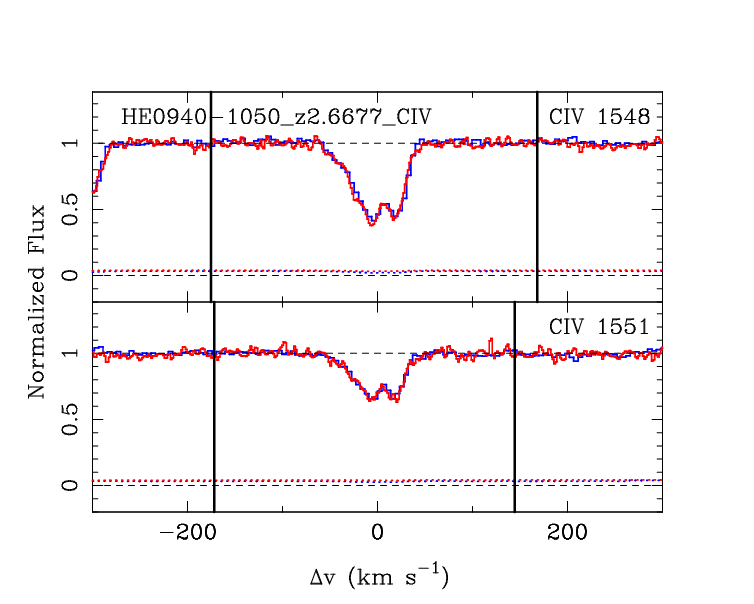}
 \end{center}
\caption{continued.}
\end{figure*}

\begin{figure*}
\addtocounter{figure}{-1}
 \begin{center}
  \includegraphics[width=8cm,angle=0]{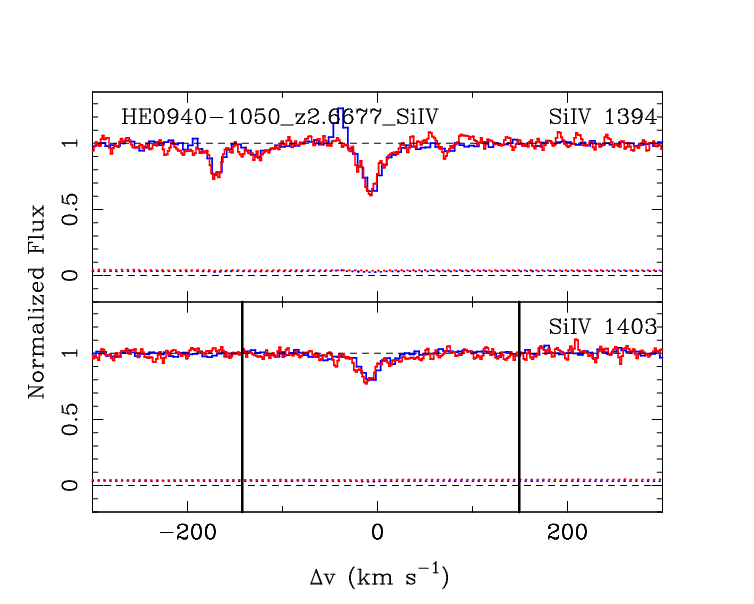}
  \includegraphics[width=8cm,angle=0]{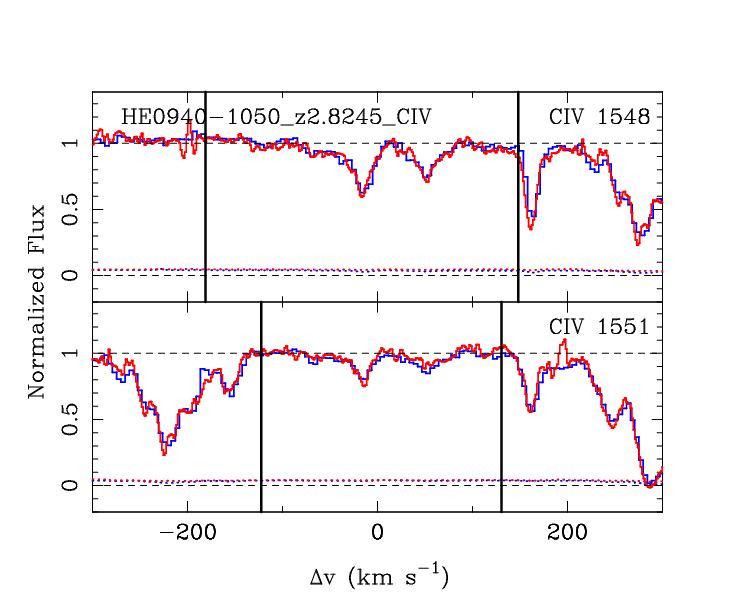}
  \includegraphics[width=8cm,angle=0]{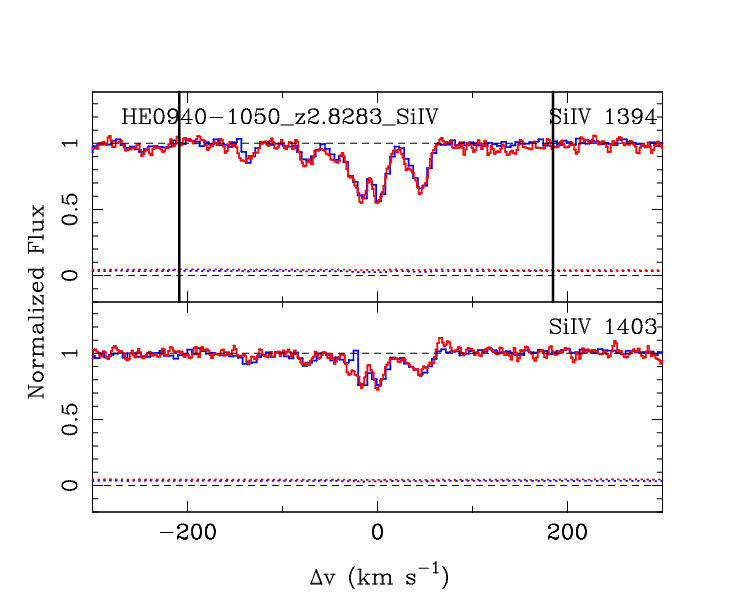}
  \includegraphics[width=16cm,angle=0]{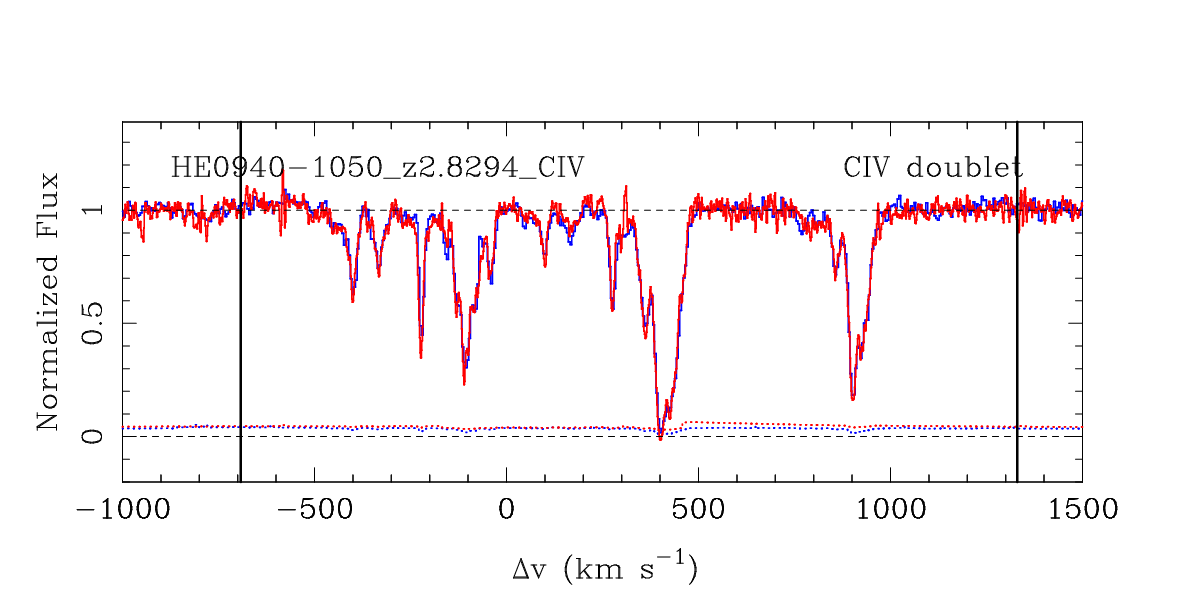}
 \end{center}
\caption{continued.}
\end{figure*}

\begin{figure*}
\addtocounter{figure}{-1}
 \begin{center}
  \includegraphics[width=8cm,angle=0]{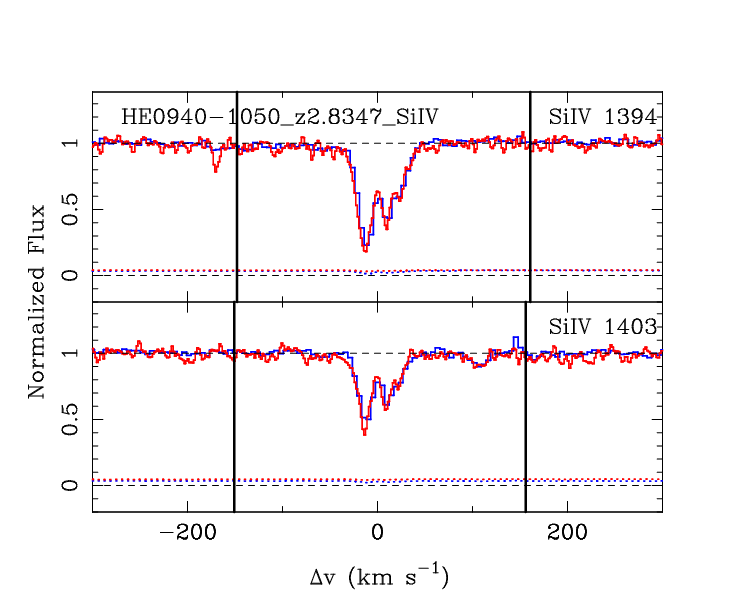}
  \includegraphics[width=8cm,angle=0]{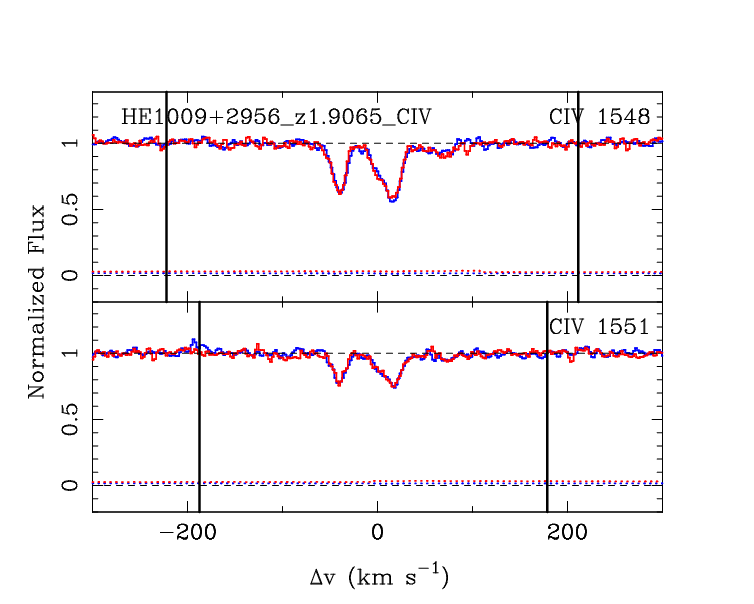}
  \includegraphics[width=8cm,angle=0]{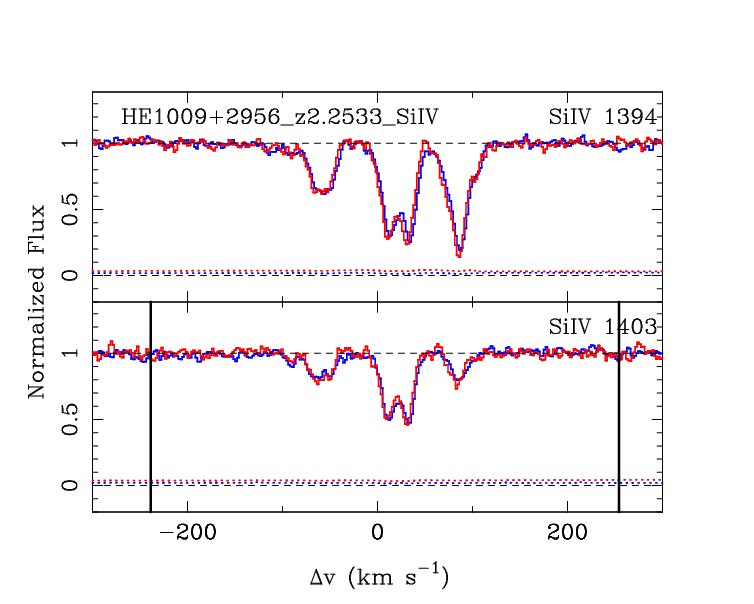}
  \includegraphics[width=8cm,angle=0]{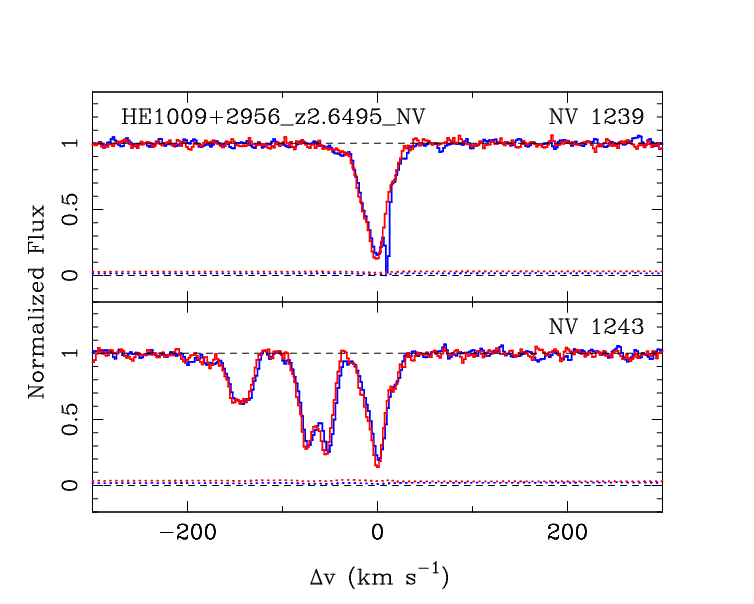}
  \includegraphics[width=8cm,angle=0]{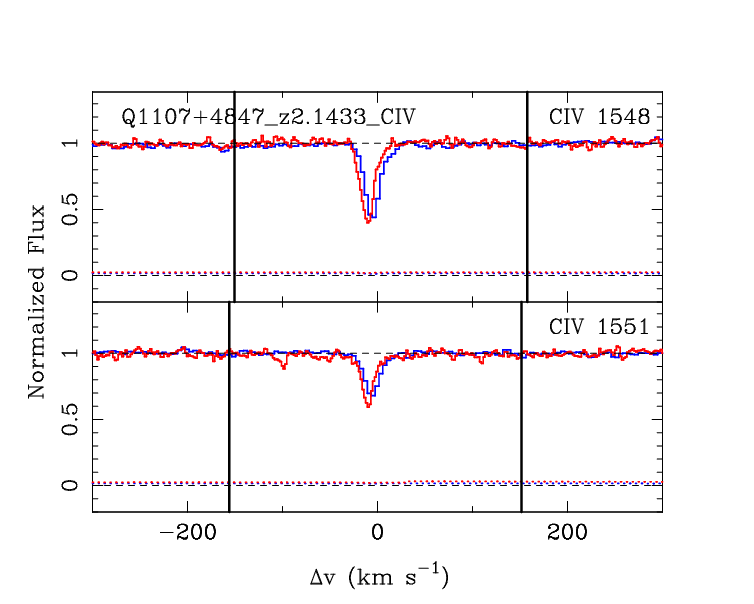}
  \includegraphics[width=8cm,angle=0]{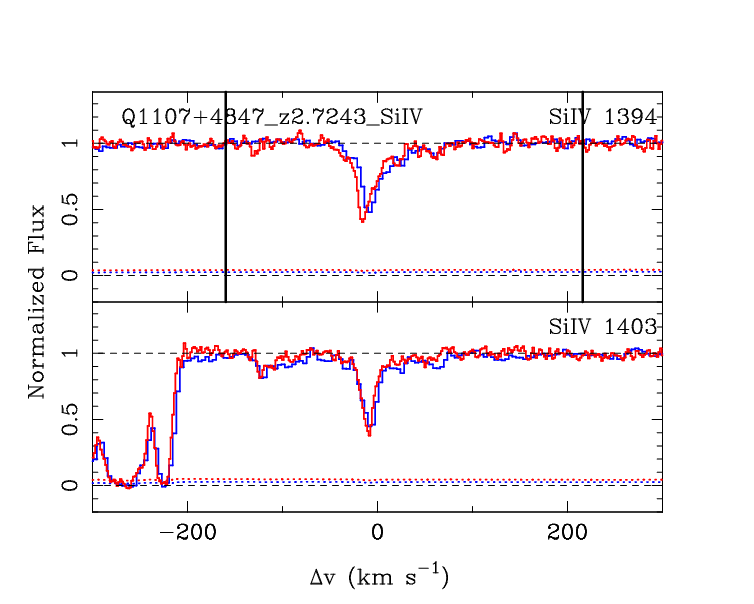}
 \end{center}
\caption{continued.}
\end{figure*}

\begin{figure*}
\addtocounter{figure}{-1}
 \begin{center}
  \includegraphics[width=8cm,angle=0]{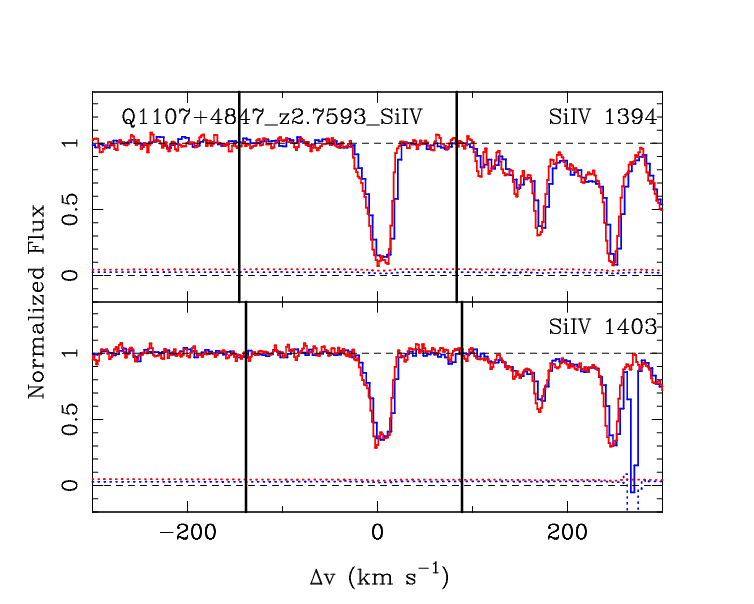}
  \includegraphics[width=8cm,angle=0]{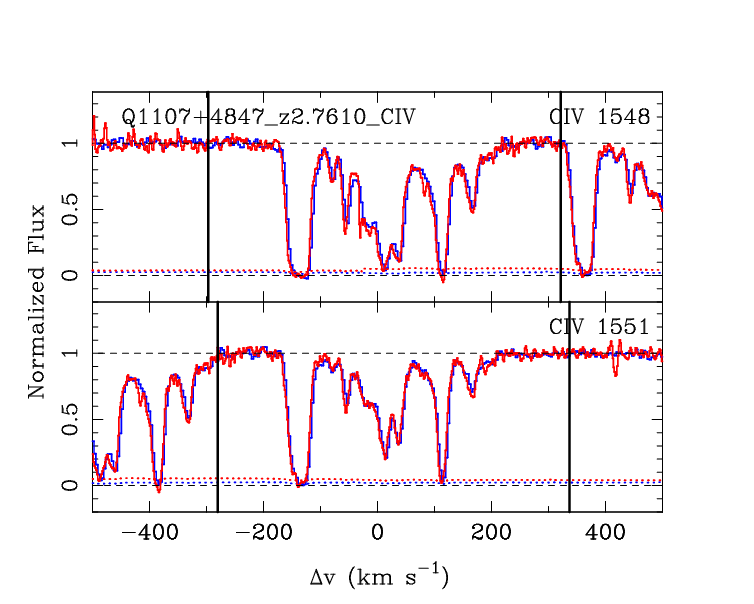}
  \includegraphics[width=8cm,angle=0]{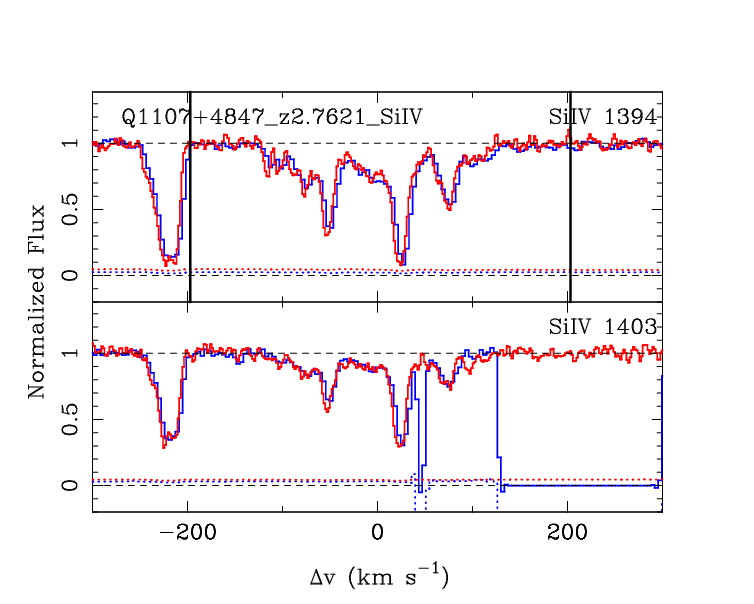}
  \includegraphics[width=8cm,angle=0]{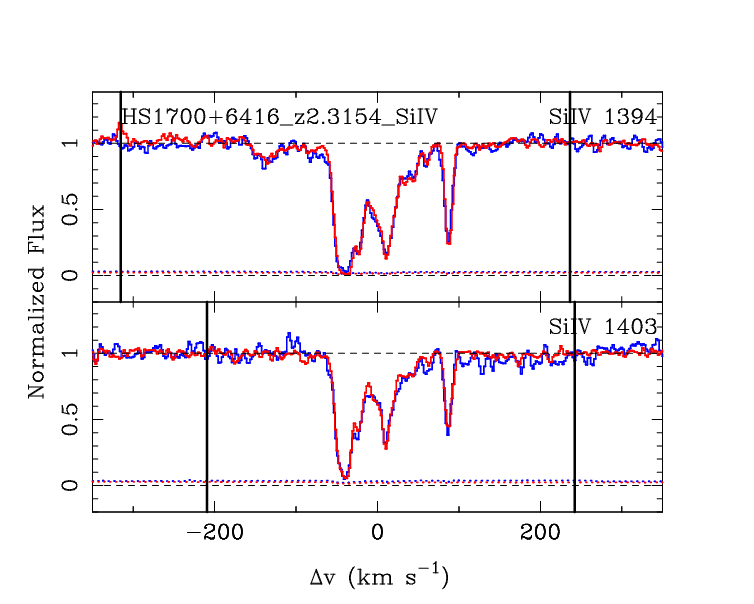}
  \includegraphics[width=8cm,angle=0]{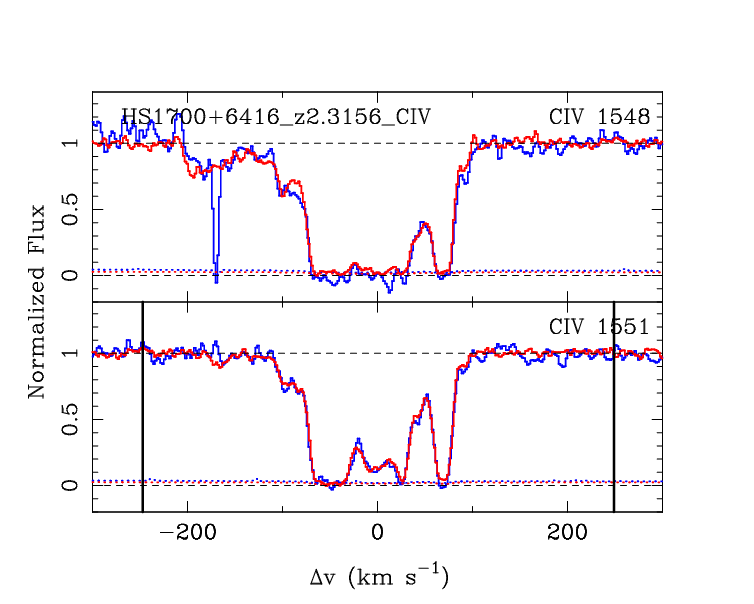}
  \includegraphics[width=8cm,angle=0]{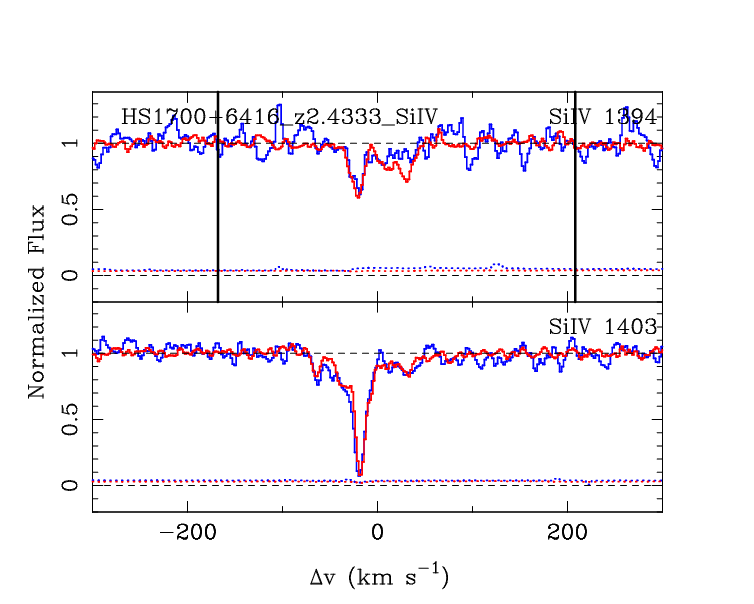}
 \end{center}
\caption{continued.}
\end{figure*}

\begin{figure*}
\addtocounter{figure}{-1}
 \begin{center}
  \includegraphics[width=16cm,angle=0]{fig1_35.ps}
  \includegraphics[width=8cm,angle=0]{fig1_36.ps}
  \includegraphics[width=8cm,angle=0]{fig1_37.ps}
  \includegraphics[width=8cm,angle=0]{fig1_38.ps}
  \includegraphics[width=8cm,angle=0]{fig1_39.ps}
 \end{center}
\caption{continued.}
\end{figure*}

\begin{figure*}
 \begin{center}
  \includegraphics[width=10cm,angle=0]{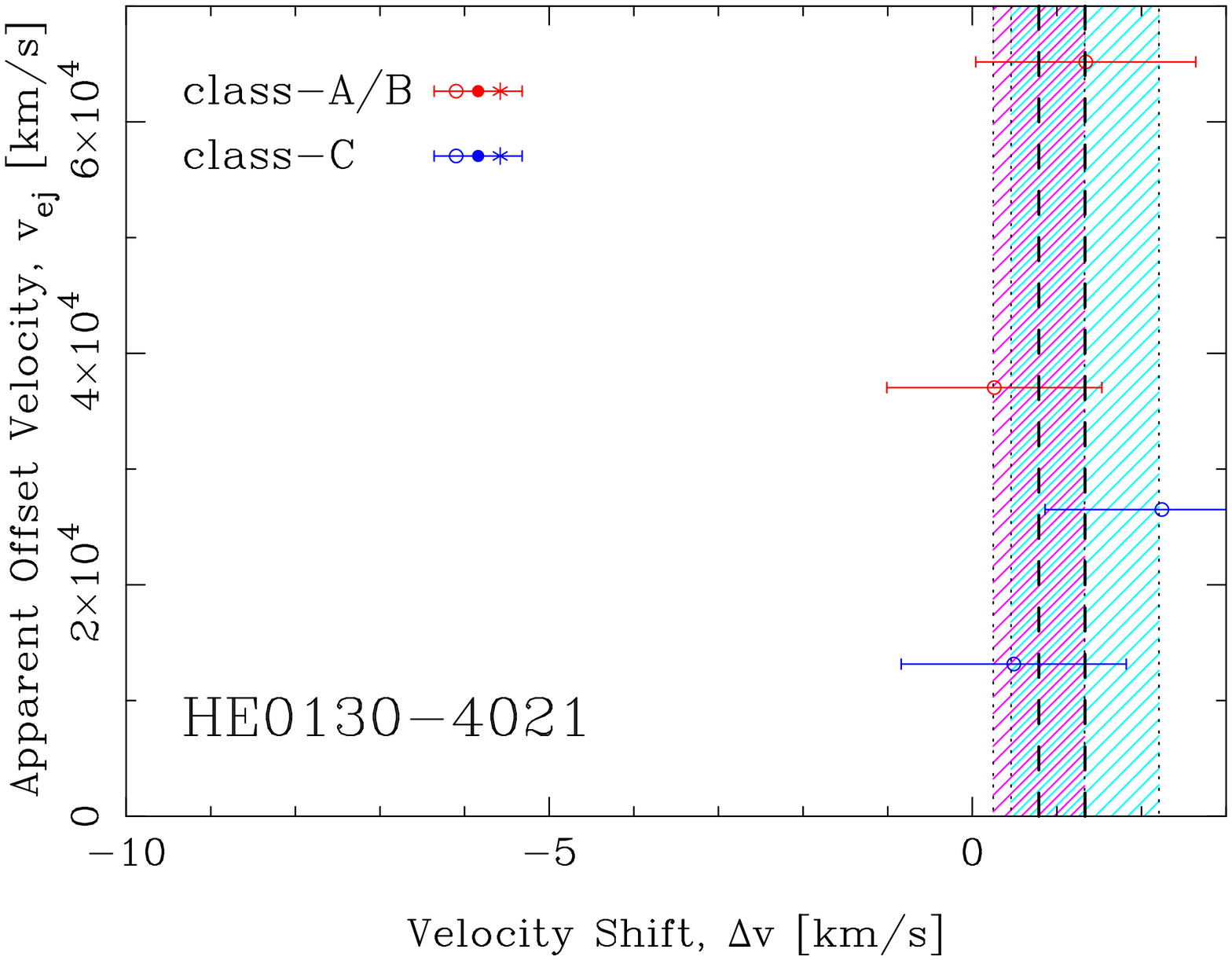}
 \end{center}
 \vspace{1cm}
 \begin{center}
  \includegraphics[width=10cm,angle=0]{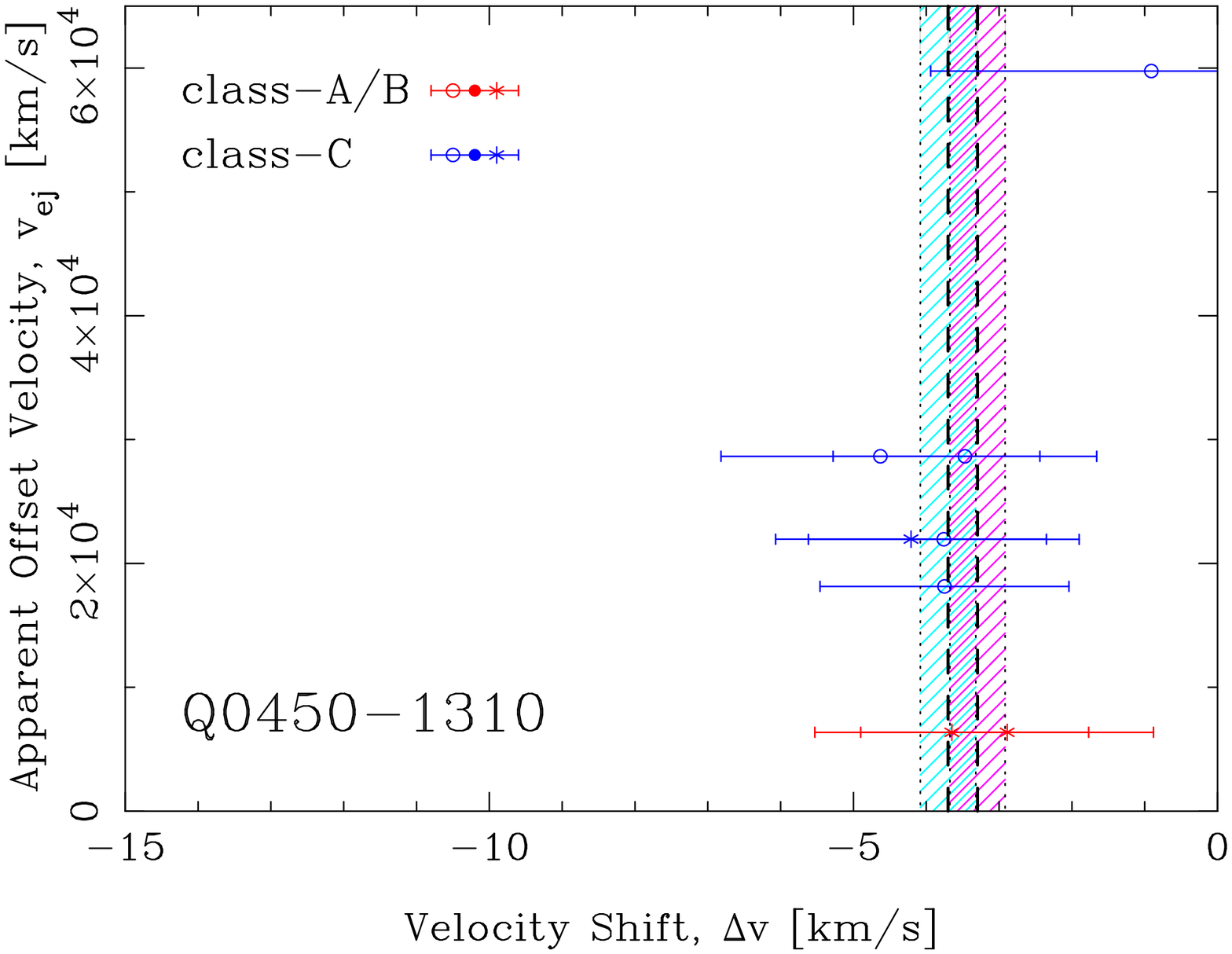}
 \end{center}
\caption{Cross-Correlation Centroid Distribution for {\it reliable}
  NALs (points with horizontal 1$\sigma$ errors) in the six quasars of
  our sample.  Open circles, filled circles, and asterisks represent
  \ion{C}{4}, \ion{N}{5}, and \ion{Si}{4} lines, respectively.  The
  horizontal axis denotes velocity shifts of absorption lines between
  observations (the second epoch relative to the first epoch), while
  the vertical axis denotes offset velocities from the quasars.  If
  both lines of a doublet are examined or if NALs from multiple
  transitions in a single absorption system are examined, multiple
  points are shown along the vertical axis at a same offset velocity.
  Hashed regions in Magenta and Cyan show the mean 1$\sigma$ errors of
  velocity shift for class-A/B and class-C NALs between observations
  that result from the combination of statistical and systematic
  errors (see Table~\ref{tab:wa}). The center and $\pm$1$\sigma$
  boundaries of velocity shifts are shown with vertical dashed and
  dotted lines.\label{fig:cccd}}
\end{figure*}

\begin{figure*}
\addtocounter{figure}{-1}
 \begin{center}
  \includegraphics[width=10cm,angle=0]{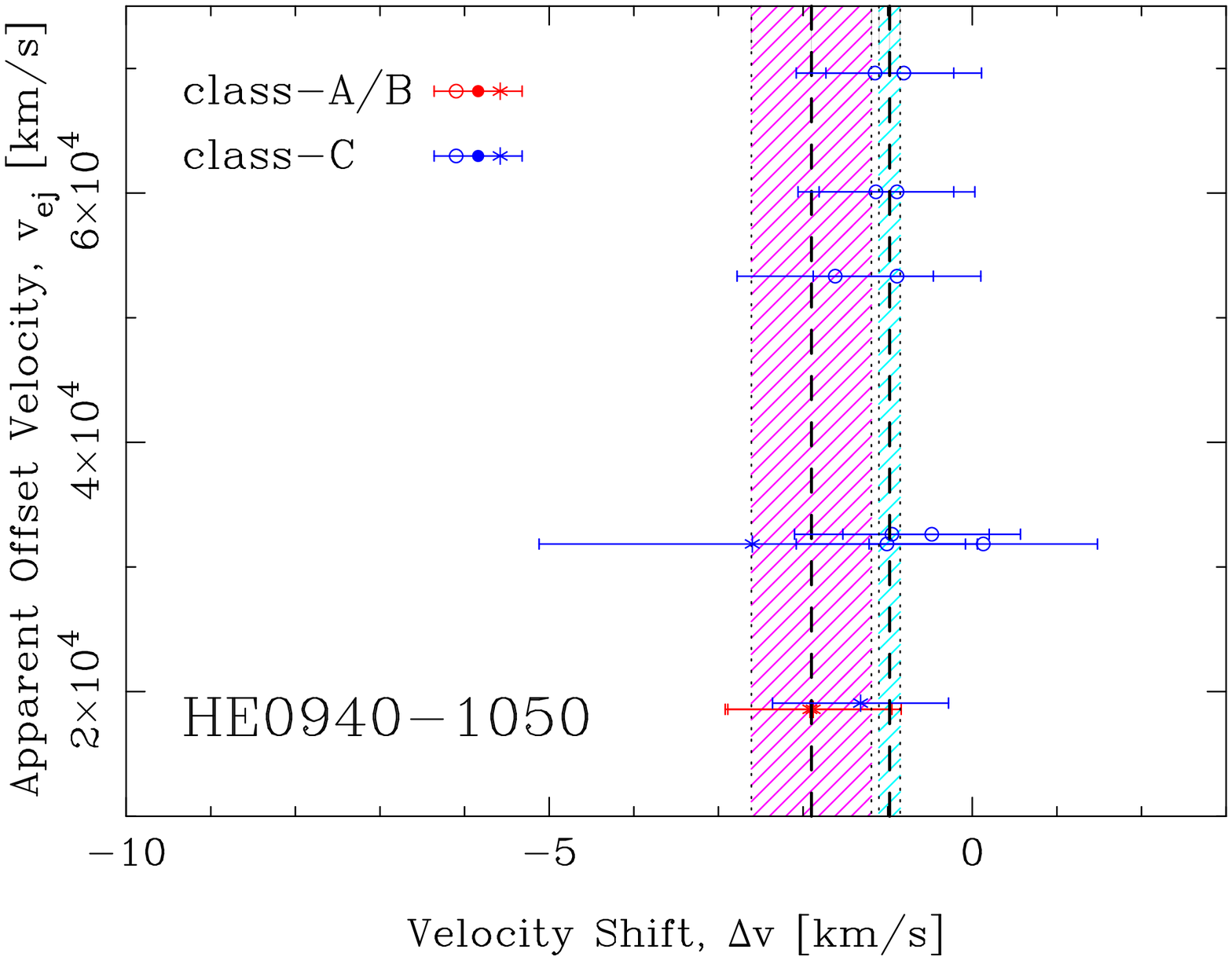}
 \end{center}
 \vspace{1cm}
 \begin{center}
  \includegraphics[width=10cm,angle=0]{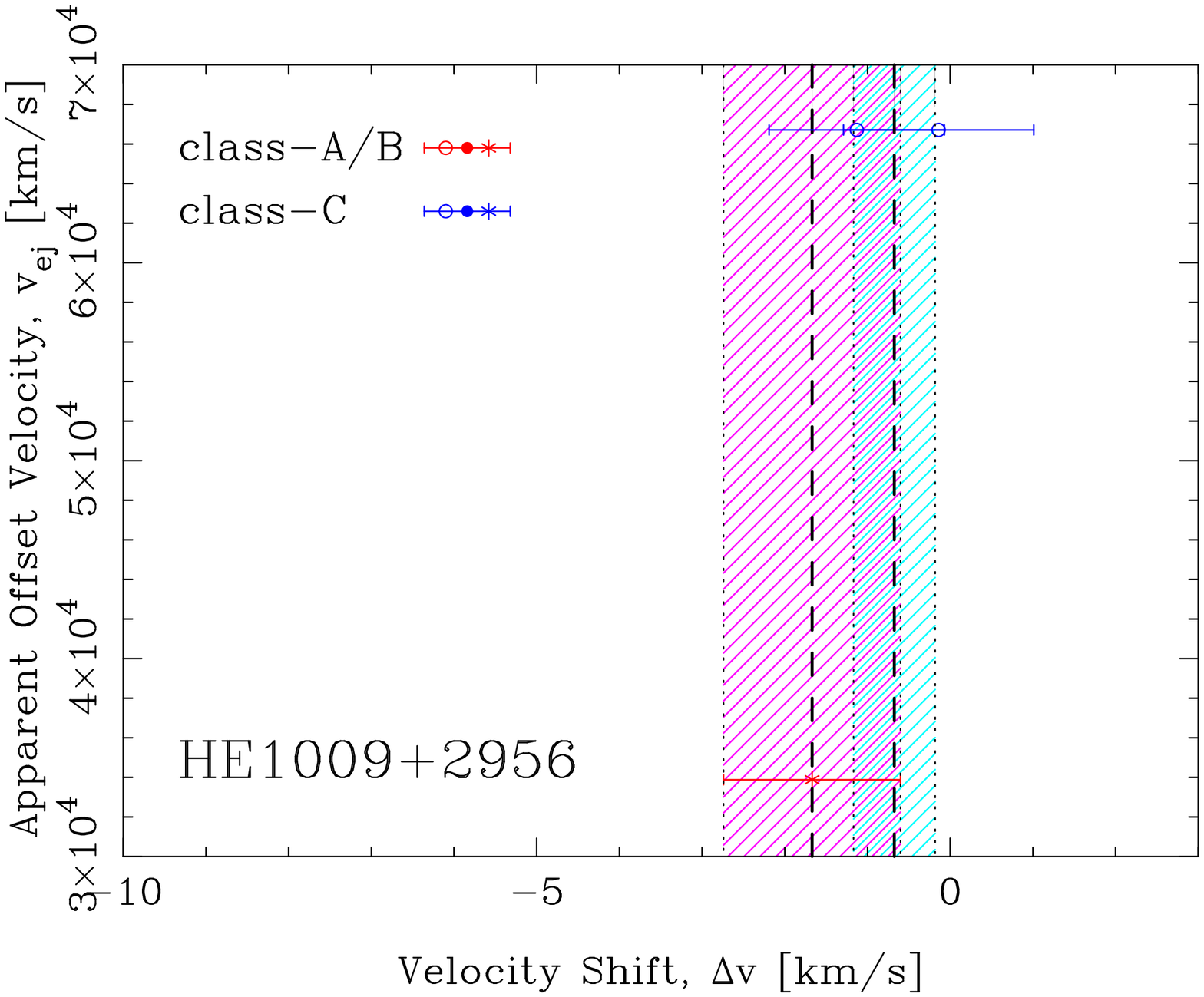}
 \end{center}
\caption{continued.}
\end{figure*}

\begin{figure*}
\addtocounter{figure}{-1}
 \begin{center}
  \includegraphics[width=10cm,angle=0]{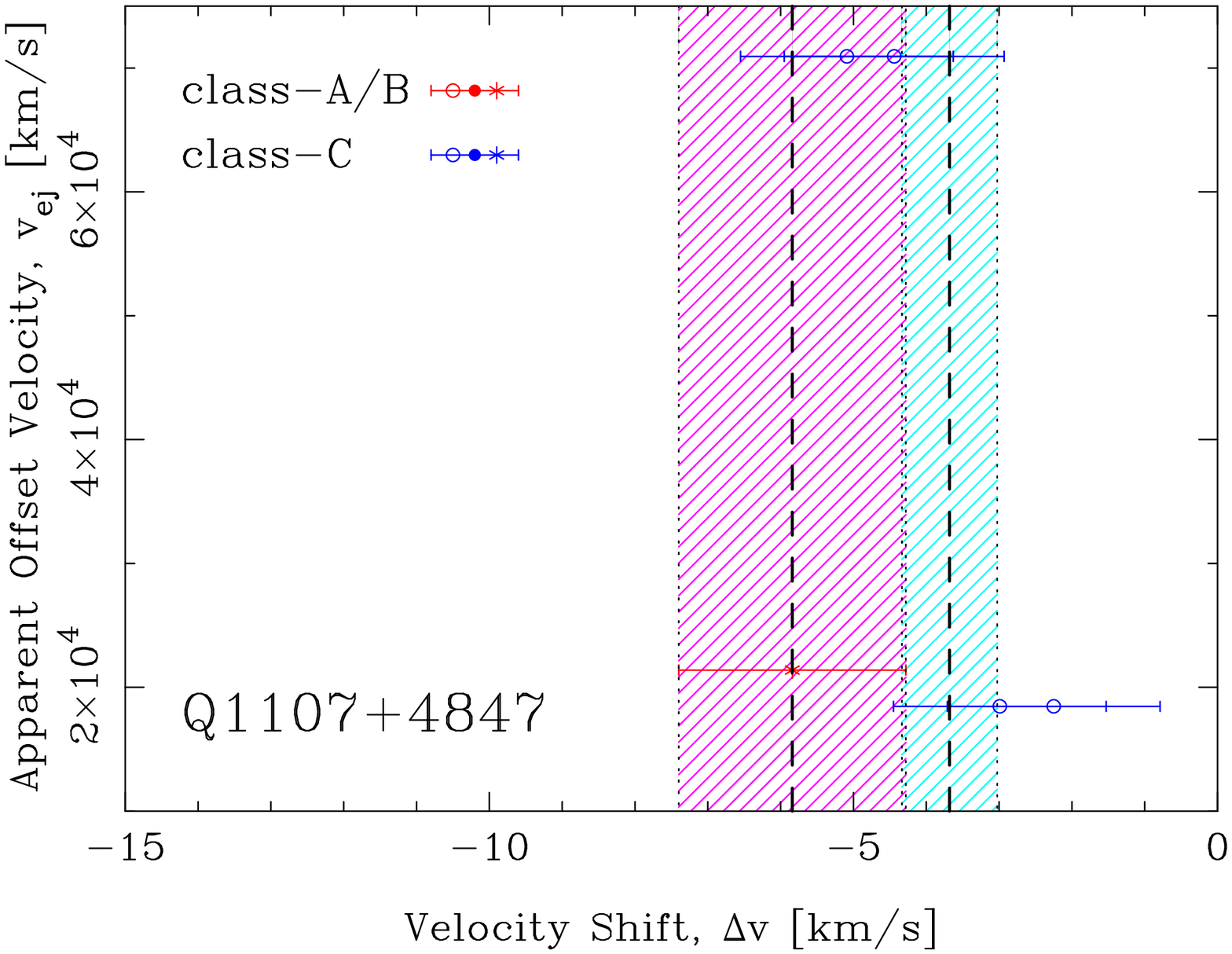}
 \end{center}
 \vspace{1cm}
 \begin{center}
  \includegraphics[width=10cm,angle=0]{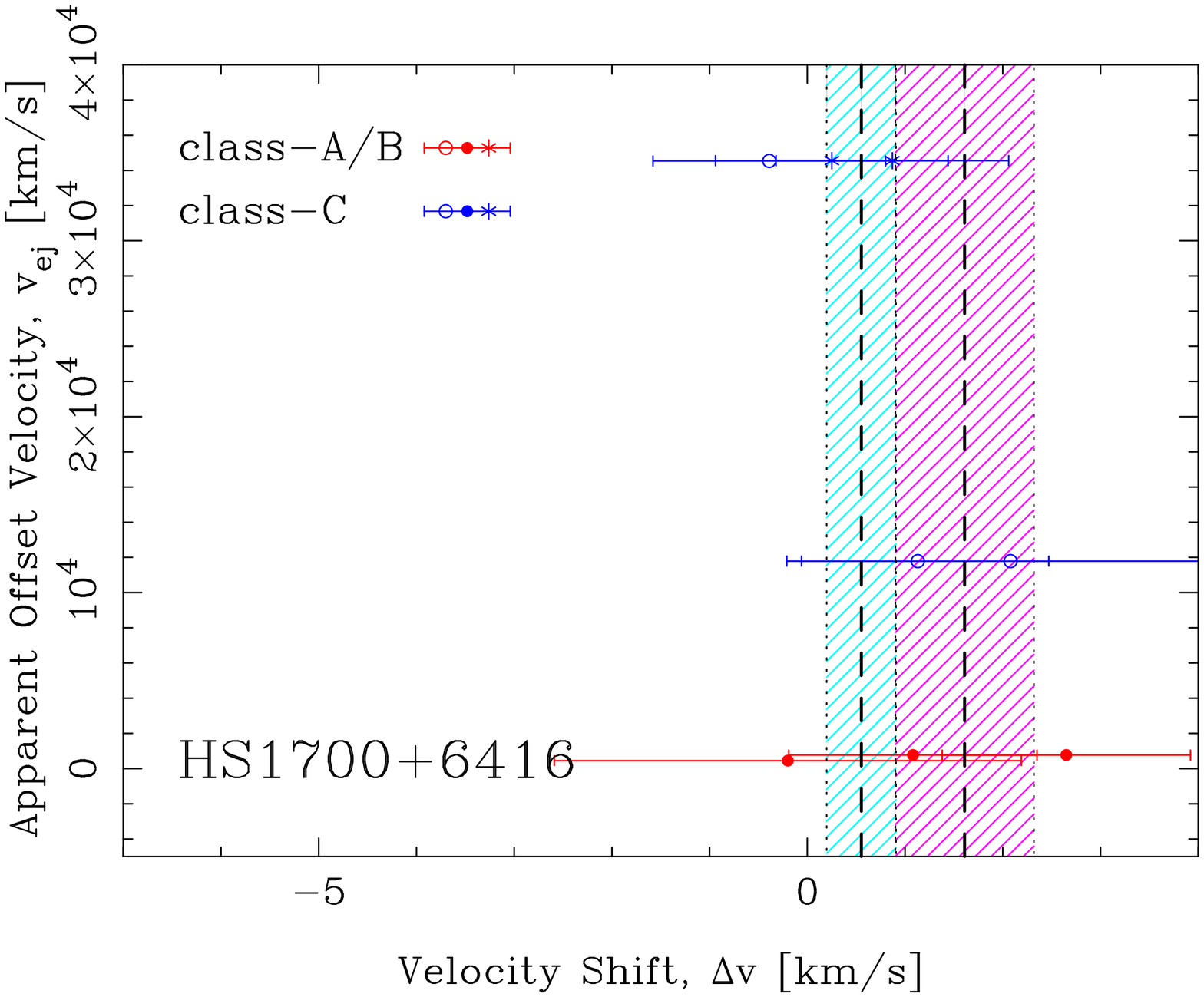}
 \end{center}
\caption{continued.}
\end{figure*}

\begin{thebibliography}{}
\bibitem[Arav et al.(2013)]{ara13} Arav, N., Borguet, B., Chamberlain,
  C., Edmonds, D., \& Danforth, C.\ 2013, \mnras, 436, 3286
\bibitem[Arav et al.(1999)]{ara99} Arav, N., Korista, K.~T., de Kool,
  M., Junkkarinen, V.~T., \& Begelman, M.~C.\ 1999, \apj, 516, 27
\bibitem[Arav(1996)]{ara96} Arav, N.\ 1996, \apj, 465, 617
\bibitem[Balbi \& Quercellini(2007)]{bal07} Balbi, A., \& Quercellini,
  C.\ 2007, \mnras, 382, 1623
\bibitem[Balsara \& Krolik(1993)]{bal93} Balsara, D.~S., \& Krolik,
  J.~H.\ 1993, \apj, 402, 109
\bibitem[Barlow \& Sargent(1997)]{bar97} Barlow, T.~A., \& Sargent,
  W.~L.~W.\ 1997, \aj, 113, 136
\bibitem[Bruni et al.(2012)]{bru12} Bruni, G., Mack, K.-H., Salerno,
  E., et al.\ 2012, \aap, 542, A13
\bibitem[Capellupo et al.(2011)]{cap11} Capellupo, D.~M., Hamann, F.,
  Shields, J.~C., Rodr{\'{\i}}guez Hidalgo, P., \& Barlow,
  T.~A.\ 2011, \mnras, 413, 908
\bibitem[Capellupo et al.(2012)]{cap12} Capellupo, D.~M., Hamann, F.,
  Shields, J.~C., Rodr{\'{\i}}guez Hidalgo, P., \& Barlow,
  T.~A.\ 2012, \mnras, 422, 3249
\bibitem[Chand et al.(2004)]{cha04} Chand, H., Srianand, R.,
  Petitjean, P., \& Aracil, B.\ 2004, \aap, 417, 853
\bibitem[Eracleous et al.(2012)]{era12} Eracleous, M., Boroson, T. A.,
  Halpern, J. P., \& Liu, J. 2012, \apjs, 201, 23
\bibitem[Evans et al.(2014)]{eva14} Evans, T.~M., Murphy, M.~T.,
  Whitmore, J.~B., et al.\ 2014, \mnras, 445, 128
\bibitem[Everett(2005)]{eve05} Everett, J.~E.\ 2005, \apj, 631, 689
\bibitem[Filiz Ak et al.(2012)]{fil12} Filiz Ak, N., Brandt, W.~N.,
  Hall, P.~B., et al.\ 2012, \apj, 757, 114
\bibitem[Fukumura et al.(2010)]{fuk10} Fukumura, K., Kazanas, D.,
  Contopoulos, I., \& Behar, E.\ 2010, \apjl, 723, L228
\bibitem[Gabel et al.(2003)]{gab03} Gabel, J.~R., Crenshaw, D.~M.,
  Kraemer, S.~B., et al.\ 2003, \apj, 595, 120
\bibitem[Gallagher et al.(2004)]{gal04} Gallagher, S.~C., Brandt,
  W.~N., Wills, B.~J., et al.\ 2004, \apj, 603, 425
\bibitem[Gibson et al.(2008)]{gib08} Gibson, R.~R., Brandt, W.~N.,
  Schneider, D.~P., \& Gallagher, S.~C.\ 2008, \apj, 675, 985-1001
\bibitem[Grier et al.(2016)]{gri16} Grier, C.~J.,
  Brandt, W.~N., Hall, P.~B., et al.\ 2016, \apj, 824, 130
\bibitem[Griest et al.(2010)]{gri10} Griest, K., Whitmore, J.~B.,
  Wolfe, A.~M., et al.\ 2010, \apj, 708, 158
\bibitem[Hall et al.(2013)]{hal13} Hall, P.~B., Brandt, W.~N.,
  Petitjean, P., et al.\ 2013, \mnras, 434, 222
\bibitem[Hall et al.(2007)]{hal07} Hall, P.~B., Sadavoy, S.~I.,
  Hutsemekers, D., Everett, J.~E., \& Rafiee, A.\ 2007, \apj, 665, 174
\bibitem[Hall et al.(2002)]{hal02} Hall, P.~B., Anderson, S.~F.,
  Strauss, M.~A., et al.\ 2002, \apjs, 141, 267
\bibitem[Hamann(1998)]{ham98} Hamann, F.\ 1998, \apj, 500, 798
\bibitem[Hamann et al.(2008)]{ham08} Hamann, F., Kaplan, K.~F.,
  Rodr{\'{\i}}guez Hidalgo, P., Prochaska, J.~X., \& Herbert-Fort,
  S.\ 2008, \mnras, 391, L39
\bibitem[Hamann et al.(2012)]{ham12} Hamann, F., Simon, L., Rodriguez
  Hidalgo, P., \& Capellupo, D.\ 2012, AGN Winds in Charleston, 460,
  47
\bibitem[Joshi et al.(2018)]{jos18} Joshi, R., Srianand, R., Chand,
  H., et al.\ 2018, arXiv:1808.05622
\bibitem[Joshi et al.(2014)]{jos14} Joshi, R., Chand, H., Srianand,
  R., \& Majumdar, J.\ 2014, \mnras, 442, 862
\bibitem[Kellermann et al.(1989)]{kel89} Kellermann, K.I., Sramek, R.,
  Schmidt, M., Shaffer, D.B., \& Green, R., 1989, \aj, 98, 119
\bibitem[Kellermann et al.(1994)]{kel94} Kellermann, K.I., Sramek,
  R.A., Schmidt, M., Green, R.F., \& Shaffer, D.B., 1994, \aj, 108,
  1163
\bibitem[Leighly et al.(2014)]{lei14} Leighly, K.~M., Terndrup, D.~M.,
  Baron, E., et al.\ 2014, \apj, 788, 123
\bibitem[L\'{\i}pari \& Terlevich(2006)]{lip06} L\'{\i}pari, S. L., \&
  Terlevich, R. J. 2006, \mnras, 368, 1001
\bibitem[Liske et al.(2008)]{lis08} Liske, J., Grazian, A., Vanzella,
  E., et al.\ 2008, \mnras, 386, 1192
\bibitem[Loeb(1998)]{loe98} Loeb, A.\ 1998, \apjl, 499, L111
\bibitem[Misawa et al.(2007a)]{mis07a} Misawa, T., Charlton, J.~C.,
  Eracleous, M., et al.\ 2007, \apjs, 171, 1
\bibitem[Misawa et al.(2007b)]{mis07b} Misawa, T., Eracleous, M.,
  Charlton, J.~C., \& Kashikawa, N.\ 2007, \apj, 660, 152
\bibitem[Misawa et al.(2014)]{mis14} Misawa, T., Charlton, J.~C., \&
  Eracleous, M.\ 2014, \apj, 792, 77
\bibitem[Murphy et al.(2003)]{mur03} Murphy, M.~T., Webb, J.~K., \&
  Flambaum, V.~V.\ 2003, \mnras, 345, 609
\bibitem[Murray et al.(1995)]{mur95} Murray, N., Chiang, J., Grossman,
  S.~A., \& Voit, G.~M.\ 1995, \apj, 451, 498
\bibitem[Proga et al.(2000)]{pro00} Proga, D., Stone, J.~M., \&
  Kallman, T.~R.\ 2000, \apj, 543, 686
\bibitem[Richards et al.(2006)]{ric06} Richards, G.~T., Lacy, M.,
  Storrie-Lombardi, L.~J., et al.\ 2006, \apjs, 166, 470
\bibitem[Rupke et al.(2002)]{rup02} Rupke, D.~S., Veilleux, S., \&
  Sanders, D.~B.\ 2002, \apj, 570, 588
\bibitem[Sandage(1962)]{san62} Sandage, A.\ 1962, \apj, 136, 319
\bibitem[Springel et al.(2005)]{spr05} Springel, V., Di Matteo, T., \&
  Hernquist, L.\ 2005, \apjl, 620, L79
\bibitem[Srianand et al.(2004)]{sri04} Srianand, R., Chand, H.,
  Petitjean, P., \& Aracil, B.\ 2004, Physical Review Letters, 92,
  121302
\bibitem[Tombesi et al.(2013)]{tom13} Tombesi, F., Cappi, M., Reeves,
  J.~N., et al.\ 2013, \mnras, 430, 1102
\bibitem[Vilkoviskij \& Irwin(2001)]{vil01} Vilkoviskij, E.~Y., \&
  Irwin, M.~J.\ 2001, \mnras, 321, 4
\bibitem[Vivek et al.(2012)]{viv12} Vivek, M., Srianand, R., Mahabal,
  A., \& Kuriakose, V.~C.\ 2012, \mnras, 421, L107
\bibitem[Wampler et al.(1995)]{wam95} Wampler, E.~J., Chugai, N.~N.,
  \& Petitjean, P.\ 1995, \apj, 443, 586
\bibitem[Webb et al.(2011)]{web11} Webb, J.~K., King, J.~A., Murphy,
  M.~T., et al.\ 2011, Physical Review Letters, 107, 191101
\bibitem[Weymann et al.(1997)]{wey97} Weymann, R.~J., Morris, S.~L.,
  Gray, M.~E., \& Hutchings, J.~B.\ 1997, \apj, 483, 717
\bibitem[Weymann et al.(1991)]{wey91} Weymann, R.~J., Morris, S.~L.,
  Foltz, C.~B., \& Hewett, P.~C.\ 1991, \apj, 373, 23
\bibitem[Wu et al.(2010)]{wu10} Wu, J., Charlton, J.~C., Misawa, T.,
  Eracleous, M., \& Ganguly, R.\ 2010, \apj, 722, 997
\end{thebibliography}
\end{document}